\documentclass[preprint2]{aastex}
\usepackage{spr-astr-addons}
\usepackage{url}
\bibpunct{(}{)}{,}{n}{,}{,}

\RequirePackage{color}

\begin{document}

 \title{Gravitational interactions between globular and open clusters: 
        an introduction}

 \shorttitle{Gravitational interactions between star clusters}
 \shortauthors{de la Fuente Marcos et al.}

 \author{R.~de~la~Fuente Marcos} 
  \and 
 \author{C.~de~la~Fuente Marcos} 
 \affil{Universidad Complutense de Madrid, Madrid, Spain} 
  \and 
 \author{D.~Reilly} 
 \affil{Suffolk University, Boston, MA, USA} 
 \email{nbplanet@fis.ucm.es}

 \begin{abstract}
    Historically, it has been assumed that globular and open clusters 
    never interact. However, recent evidence suggests that: globular 
    clusters passing through the disk may be able to perturb giant 
    molecular clouds (GMCs) triggering formation of open clusters and 
    some old open clusters may be linked to accreted globulars. Here, 
    we further explore the existence of possible dynamical connections 
    between globular and open clusters, and realize that the most 
    obvious link must be in the form of gravitational interactions. If 
    open clusters are born out of GMCs, they have to move in similar 
    orbits. If we accept that globulars can interact with GMCs, 
    triggering star formation, it follows that globular and open 
    clusters must also interact. Consistently, theoretical arguments 
    as well as observational evidence, show that globular and open 
    clusters certainly are interacting populations and their 
    interactions are far more common than usually thought, especially 
    for objects part of the bulge/disk. Monte Carlo calculations 
    confirm that conclusion. Globular clusters seem capable of not 
    only inducing formation of open clusters but, more often, their 
    demise. Relatively frequent high speed cluster encounters or 
    cluster harassment may also cause, on the long-term, slow erosion 
    and tidal truncation on the globulars involved. The disputed 
    object FSR 1767 (2MASS-GC04) may be, statistically speaking, the 
    best example of an ongoing interaction. 
 \end{abstract}


 \section{Introduction}
    It is becoming increasingly clear that the observed field stars are mostly the by-product of the disruption of some type of 
    stellar ensemble (see e.g. Hopkins 2013), bound (star cluster) or unbound (stellar association). The vast majority of clusters 
    or stellar aggregates are not long-term stable and dissociate into individual stars shortly after their formation (e.g. 
    Tutukov 1978; de la Fuente Marcos and de la Fuente Marcos 2004; de Grijs 2009; Goodwin 2009); moreover, the majority of stars 
    appear to form in low density environments --associations-- not dense star clusters (e.g. Fritze 2009; Bressert et al. 2010; 
    Gieles and Portegies Zwart 2011). The fraction of all stars in the Universe once formed in bound star clusters is currently 
    estimated at 30--35~\% (Kruijssen 2012). These facts explain why, at any epoch, most stars in a galaxy are not associated to 
    clusters but to the overall field. Within galaxies, a range of external forces can trigger giant molecular cloud (GMC) 
    fragmentation which in turn, leads to stellar association and star cluster formation within star-forming complexes (see e.g. 
    Elmegreen and Lada 1977; Efremov 1978, 1979; Elmegreen and Efremov 1996; Efremov and Elmegreen 1998). In this context, star 
    clusters and associations appear as primary galactic building blocks. 

    Historically, star clusters in the Milky Way are split up into two distinct, fully independent, seemingly unrelated 
    populations: globular and open clusters. In general terms, this traditional view appears to be well supported by available 
    observational data: globular and open clusters show significantly different structural and kinematic properties and the two 
    groups seem to be of rather different origin (see e.g. Sparke and Gallagher 2007). Consistently, it has been customarily 
    assumed that globular and open clusters never interact; i.e., there are no connections between the two types of clusters. 
    However, this classical and prevalent interpretation is now open to question, at least partially, because recent evidence 
    appears to indicate that: (i) some old open clusters may be linked to accreted globular clusters (see e.g. Carraro and Bensby 
    2009) and (ii) globular clusters passing through the disk may trigger formation of open clusters (see e.g. Vande Putte and 
    Cropper 2009). These certainly are interesting scenarios but the novelty of these ideas also raises questions about their 
    actual feasibility and proper characterization. For example and based on currently available data, how strong is the 
    statistical evidence for these globular-open cluster connections? and more importantly, what criteria should be used to define 
    such connections, if real? In the first case, Carraro and Bensby (2009) suggest that globular and open clusters that were 
    accreted together are expected to have similar chemical and kinematic properties (although the need of a common chemical 
    signature is a debatable point). For the second scenario, an excess of young open clusters close to a globular cluster may 
    hint at a potential cause-and-effect relationship. In this case, no kinematic link is expected: why should an open cluster 
    formed out of a shocked GMC have a motion similar to that of a globular cluster passing through the Galactic disk? Both, 
    globular cluster and molecular cloud will have their orbits perturbed by the interaction, but not enough to give the newly 
    formed open clusters and their progenitor a common motion. No chemical connection is expected, either. In this case, only 
    positional evidence in the form of an excess of small pair separations for some objects can be used to argue for a possible 
    parental connection.  
    
    The particular cases pointed out above are all part of a much more general problem: how often are globular and open clusters 
    interacting, even if weakly, in an environment similar to the Solar Neighbourhood? Are these putative interactions relatively 
    frequent or, on the contrary, very rare events? If frequent, what is their most likely outcome? Here, we attempt to provide an 
    answer to these rather general questions; but, we also search for specific globular clusters in close proximity to a larger 
    than average number of young open clusters and explore the possibility that some old open clusters may be linked to accreted 
    globular clusters. Before going into any further details, we must also point out that the existence of gravitational 
    interactions between globular and open clusters is, in fact, both intuitive and obvious the moment we accept that globular 
    clusters and GMCs interact. Open clusters are born out of GMCs; therefore, they must follow very similar orbits and their 
    kinematics must be virtually equivalent, especially for the youngest open clusters. If we admit that globular clusters can 
    interact with GMCs, it immediately follows that globular and open clusters must be able to interact as well. For interacting 
    globular clusters and GMCs we witness creation of stars; for interacting globular and open clusters the most obvious outcome 
    should be tidal disruption of the open cluster. It is true that GMCs are more massive and larger in size than open clusters 
    but open clusters both outnumber and outlive GMCs. In this context, neglecting the existence of globular-open cluster 
    interactions is equivalent to negate the obvious. The widely accepted assumption that globular and open clusters never 
    interact probably has its roots in the fact that no globular clusters appear to exist within 1 kpc from the Sun, the only 
    region of the Milky Way where our open cluster census is reasonably complete. This combination of circumstances induces an 
    obvious bias that conditions our approach to the general problem studied here. For this reason, our paper is organized as 
    follows. The theoretical, expected frequency of globular-open cluster interactions is studied in Sect. 2 using Kinetic Theory. 
    The expectations associated to a non-collisional scenario are obtained in Sect. 3 using both analytical results and Monte 
    Carlo techniques. The issue of GMC-star cluster interactions as a competing/cooperating process is discussed in Sect. 4. 
    Observational data are introduced and an initial cluster separation analysis is performed in Sect. 5. The statistical 
    significance of our findings is studied in Sect. 6. A discussion and a Monte Carlo-based comparison between theoretical 
    expectations and actual observational evidence are presented in Sect. 7. In Sect. 8, we focus on an apparent outlier: FSR 
    1767 (2MASS-GC04). Our results are placed within the context of other recent, related studies in Sect. 9. Section 10 
    summarizes our conclusions.

 \section{Globular-open cluster interactions}
    The currently available information on star clusters in the Milky Way is incomplete and likely biased (see below); if we want 
    to find out how often globular and open clusters are expected to interact in the Solar Neighbourhood, we should make few a 
    priori assumptions and use theoretical arguments starting from general physical principles. This is the only approach that can 
    yield solid conclusions when exploring the subject of globular-open cluster dynamical interactions and comparing with 
    observational data.
%
%---------------------------------------------------------------------------------------------------------------------------------
%
    \begin{figure}[!htb]
      \centering
        \includegraphics[width=\columnwidth]{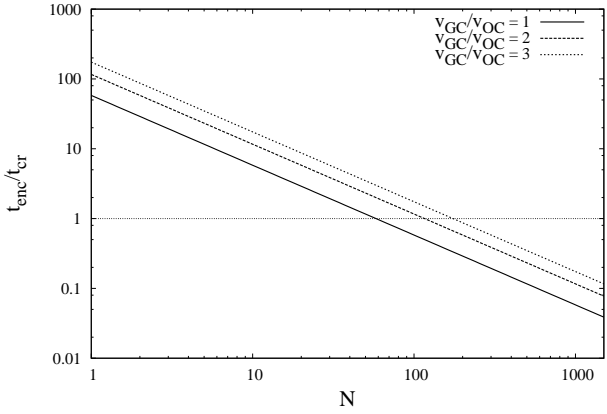}
        \caption{Evolution of $t_{\rm enc}/t_{\rm cr}$ as a function of $N$ assuming $r$ = 1 kpc, $r_t$ = 35 pc and three 
                 representative values of $v_{\rm GC}/v_{\rm OC}$. Encounters are only important if $t_{\rm enc}/t_{\rm cr} < 1$. 
                 For regions with high density of open clusters like the spiral arms or the bulge (with $N \sim 10^3$) the number 
                 of encounters per disk crossing could be as high as 50; for the typical environment of the Solar Neighbourhood 
                 could be around 10 (nearly 400 open clusters within 1 kpc, see the text for details). Globular clusters cross the
                 disk twice per orbit. This plot displays results from Eq. (\ref{timescale}) which are nearly five times larger 
                 than those from the more realistic Eq. (\ref{timescale1}).}
      \label{timescaleplot}
    \end{figure}
%
%---------------------------------------------------------------------------------------------------------------------------------
%

    In the framework of Kinetic Theory (see e.g. Binney and Tremaine 2008), let us consider a sphere of radius $r$ enclosing $N$ 
    open clusters. A globular cluster of tidal radius $r_t$ is moving across that sphere. The number density of open clusters 
    within that sphere can be written as $\rho = 3 \ N (4 \ \pi \ r^3)^{-1}$ and they move with a root mean square (RMS, the RMS 
    value is a measure of the average value) velocity $v_{\rm OC}$; $v_{\rm OC}$ is the average velocity with which open clusters 
    move within that sphere. Following Innanen et al. (1972), for two clusters separated by a distance larger than three times 
    the outer radius of each cluster, the amount of mutual disruption is rather negligible. The average value of the tidal radius 
    for open clusters in the Milky Way disk is 10 pc and for globular clusters is 35 pc (Binney and Tremaine 2008). The 
    interaction cross-section for such a globular cluster can be written as $\Sigma = \pi \ (3 \ r_t)^2$. The characteristic 
    time-scale on which a globular cluster experiences an encounter with an open cluster is given by $t_{\rm enc} \approx 
    (\rho \ \Sigma \ v_{\rm OC})^{-1}$. In terms of the time taken by the globular cluster to cross the sphere or crossing time 
    $t_{\rm cr} = 2 \ r \ v_{\rm GC}^{-1}$, where $v_{\rm GC}$ is the characteristic velocity of the globular cluster, we can 
    write
    \begin{equation}
       \frac{t_{\rm enc}}{t_{\rm cr}} = \frac{2}{27 \ N} \frac{v_{\rm GC}}{v_{\rm OC}}
                                         \left(\frac{r}{r_t}\right)^2 \,.
                                          \label{timescale}
    \end{equation}
    Encounters are only important if $t_{\rm enc}/t_{\rm cr} < 1$. Figure \ref{timescaleplot} displays Eq. (\ref{timescale}) as a 
    function of the number of open clusters $N$ for three representative values of the ratio $v_{\rm GC}/v_{\rm OC}$ (assumed to 
    be in the range [1, 3], see below). If a globular cluster enters an environment like the Solar Neighbourhood, with nearly 400 
    open clusters within 1 kpc (see below), some encounters are expected; besides, the probability of undergoing discrete 
    encounters reaches its peak value twice per globular cluster orbital period and globular clusters are very long lived. 
    Therefore, the probability of suffering at least one relatively close encounter with an open cluster integrated over the 
    entire life of a typical globular cluster is far from zero. It is clear that, statistically speaking, the role of these 
    interactions cannot simply be ignored.  

    At this point, it may be argued that it is often assumed that $v_{\rm GC} \gg v_{\rm OC}$. Under that (wrong) assumption, 
    $t_{\rm enc}/t_{\rm cr} \gg 1$ and the role of encounters becomes completely negligible. However, it is untrue that the 
    typical total Galactocentric velocities of globular and open clusters are so different. The local escape speed, 498 $< v_{\rm 
    esc} <$ 608 km~s$^{-1}$ (Smith et al. 2007), represents an upper limit to the total velocity of both globular and open 
    clusters; if a globular cluster is bound to the Milky Way, $v_{\rm GC} <$ 500-600 km~s$^{-1}$. On the other hand, it is 
    widely accepted that most open clusters move around the centre of the Galaxy in almost-circular orbits. The local circular 
    orbital velocity for objects in the disk is about 220 km~s$^{-1}$. These values are fully consistent with our choice of [1, 3] 
    as the range in $v_{\rm GC}/v_{\rm OC}$. In more detail and if we consider the well studied (but nonetheless incomplete) 
    sample (new Hipparcos catalogue) of 20 open clusters described in van Leeuwen (2009), the average total Galactocentric 
    velocity for open clusters in the Solar Neighbourhood is 224$\pm$7 km~s$^{-1}$. This value has been computed using the 
    Heliocentric velocity components from table 7 in van Leeuwen (2009), the peculiar motion of the Sun relative to the Local 
    Standard of Rest (LSR) in Sch\"onrich et al. (2010) and the in-plane circular motion of the LSR around the 
    Galactic centre. The RMS value of the velocity component perpendicular to the Galactic plane for this sample of 20 open 
    clusters is 4.6 km~s$^{-1}$. As for globular clusters, Kalirai et al. (2007) found that for NGC 6397 (one of the objects of 
    interest in this research, see below) the total velocity relative to the Galactic centre is 195 km~s$^{-1}$ with a vertical 
    velocity of -140 km~s$^{-1}$; this cluster has made frequent passages through the Galactic disk. Such a globular cluster has 
    a total velocity similar to that of an open cluster yet it crosses the thick disk in about 10 Myr. For the samples of 
    globular clusters in Casetti-Dinescu et al. (2007, table 3) and Casetti-Dinescu et al. (2010, table 5) we obtain an average 
    total Galactocentric velocity of 175$\pm$91 km~s$^{-1}$ and the RMS value of the velocity component perpendicular to the 
    Galactic plane is 69 km~s$^{-1}$. This group of 15 globular clusters have Galactocentric distances in the range 0.7-11 kpc 
    with separations from the Galactic plane in the range [-1.9, 2.7] kpc. The slowest vertical motion is -7 km~s$^{-1}$ for NGC 
    6284 and the largest is -146 km~s$^{-1}$ for NGC 6293. Globular clusters crossing the disk typically spend between 5 and 150 
    Myr to travel 1 kpc; this translates into 1 to 13 encounters (at least) per crossing in the most typical cases, the ones of 
    interest here.  

    So far we have neglected the fact that, in the Galaxy, open clusters are mainly found in the disk. If we include the disk's 
    cylindrical geometry, our results are similar:
    \begin{equation}
       \frac{t_{\rm enc}}{t_{{\rm cr}_1}} = \frac{1}{18 \ N} \frac{v_{\rm GC}}{v_{\rm OC}}
                                             \frac{r \ h}{r_t^2} \,,
                                              \label{timescale1}
    \end{equation}
    where we assumed a volume of $\pi \ h \ r^2$ for the section of the disk with $h$, the characteristic scale height of the 
    disk. Equation (\ref{timescale1}) assumes a globular cluster moving in an orbit with a small inclination with respect to the 
    disk. For a polar orbit (i.e., a globular cluster crossing perpendicular to the disk), the following expression is obtained:
    \begin{equation}
       \frac{t_{\rm enc}}{t_{{\rm cr}_2}} = \frac{1}{9 \ N} \frac{v_{\rm GC}}{v_{\rm OC}}
                                             \left(\frac{r}{r_t}\right)^2 \,,
                                              \label{timescale2}
    \end{equation}
    which represents the most unfavourable case regarding globular-open cluster interactions. The ratio between the two 
    expressions (\ref{timescale1} and \ref{timescale2}) is of order $h/r \sim 0.3$. The interaction rate (1/$t_{enc}$) predicted 
    by Eq. (\ref{timescale1}) is nearly 5 times larger than that from Eq. (\ref{timescale}). The actual number of 
    encounters can be even larger because our simplified analysis is neglecting gravitational focusing; i.e., the possibility 
    that two clusters initially having an impact parameter $\gg 3 \ r_t$, be brought closer together due to their mutual 
    gravitational attraction as the relative velocities during the encounter could be, in some cases, similar to the escape 
    velocity from the globular cluster. In high star cluster density environments like a spiral arm or near the Galactic centre, 
    dozens of encounters may be experienced within just 10 Myr (this is the time-scale to cross 2 kpc at about 200 km/s). In 
    sharp contrast and for the outer disk, globular-open cluster interactions are rare events. At high altitude over the Galactic 
    disk these encounters are completely negligible. In summary, a globular cluster traveling across an environment similar to 
    the Solar Neighbourhood is expected to suffer about ten interactions with resident open clusters. In the Milky Way, there are 
    about 58 globular clusters currently located $<$ 1 kpc from the Galactic plane. For those, the interaction rate is $\geq$ 1 
    Myr$^{-1}$. The two-phase system made of globular and open clusters is far from collisionless. 

    But, what is the typical outcome to be expected after one of these encounters? When two star clusters undergo a close 
    encounter we may observe the formation of a transient binary system ending in merging (if the relative velocity is low enough 
    and both clusters have similar masses), full destruction of the less massive cluster, or (more often) a hyperbolic encounter 
    in which both clusters emerge relatively unaffected and eventually separate (de la Fuente Marcos and de la Fuente Marcos 2010). 
    Most of the encounters are likely to be distant and may not have a major impact on the dynamics of the open cluster involved 
    but if the minimum impact parameter is close to $r_t$, the open cluster will be fully destroyed in a very short time-scale 
    (see e.g. figs 8 and 9 in de la Fuente Marcos and de la Fuente Marcos 2010). As for the effects on the globular clusters 
    themselves, the continuous succession of even weak high speed encounters must cause slow erosion and eventually produce severe 
    truncation by tidal forces. The mechanism is conceptually similar to the galaxy harassment, first described by Moore et al. 
    (1996), in galaxy clusters. Repeated encounters can also induce a very low value for the luminosity of the affected clusters 
    due to the loss of stars resulting from tidal effects. On the other hand, low-mass stars, which are more likely to be found in 
    the outer regions of star clusters because of mass segregation, will be stripped off preferentially. Encounters gradually 
    perturb stars away from the path that they would have followed in a strictly collisionless environment. The characteristic 
    time over which the loss of dynamical memory occurs is called the relaxation time. The actual impact of the encounters on the 
    clusters can be better quantified using the relaxation time of the system that is $\gtrsim$ 100 Myr for globular clusters and 
    $\lesssim$10 Myr (with a range of 5-200 Myr) for typical open clusters. In both cases, the effects of the encounters are not 
    negligible as the relaxation time is longer than the characteristic time between encounters. However, globular clusters can 
    better recover as many of them spend just about 20 Myr per orbital period moving in regions where encounters are possible. The 
    relaxation time is longer far from the centre of the cluster and shorter at the cluster core, the central regions of the 
    clusters may be able to survive the frequent encounters but the outer regions will gradually be lost into the field. The 
    picture that emerges from our analysis contradicts the conventional assumption that globular and open clusters are completely 
    unrelated populations and never interact. The globular-open cluster dynamical connection uncovered here should translate into 
    several observable trends and features among clusters within a few kpc from the Sun:
    \begin{itemize}
       \item The number of globular-open cluster pairs with separations under $\sim$105 pc (three times the average tidal radius 
             for globulars in the Milky Way) must be very small as open clusters interacting with a typical globular cluster at 
             such short distance will not survive for long. 
       \item The number of globular-open cluster pairs should decrease with the separation, likely as a power-law.
       \item Globular clusters within the disk should be surrounded by a relatively large number of open clusters but few of them 
             will be old (age $>$ 1 Gyr) objects.
       \item Globular clusters that have been involved in interactions should exhibit some level of tidal truncation. 
       \item The effects of this dynamical connection should be more important towards the bulge and the Galactic centre.
       \item Open clusters following highly inclined orbits must be well protected against the effects of the encounters described 
             here. This should translate into an obvious difference between the relative number of old open clusters observed at 
             high and low Galactic latitude (more properly, altitude).
    \end{itemize} 
    The cluster harassment described here is just another mechanism to induce tidal truncation in globular clusters and operates 
    concurrently with disk and bulge shocking as described by (e.g.) Chernoff et al. (1986) or Fall and Zhang (2001) and 
    identified observationally by Leon et al. (2000). Cluster harassment has recently been discussed within the context of the 
    disruption of star clusters in a hierarchical interstellar medium (Elmegreen and Hunter 2010). A tidally truncated globular 
    cluster may still have a relatively large radius if it hosts a sizable population of black holes (Strader et al. 2012); in
    addition, the average radius of globular clusters increases with increasing Galactocentric distance. Our theoretical analysis 
    also indicates that interactions between globular clusters are extremely rare and for all practical purposes we can assume 
    that globulars do not interact with other globular clusters in today's Milky Way. As for open clusters orbiting globular 
    clusters within the Milky Way, the dynamical scenario presented here makes that situation highly unlikely. Only a very 
    recently accreted globular cluster (more properly, dwarf galaxy) may still be able to retain its own distinctive cohort of 
    open clusters.

 \section{A purely non-collisional scenario}
    The issue of the existence of a dynamical connection between globular and open clusters has been largely ignored in Galactic 
    studies; contrary to our findings above, it is customarily assumed that globular and open clusters never interact. However, 
    if such non-collisional scenario is valid, the distribution of intercluster distances should exhibit a distinctive shape. In 
    this section, we focus on the geometrical consequences of neglecting gravitational effects and assume that globular and open 
    clusters never interact. Globular clusters are mainly found towards the Galactic centre and open clusters are mainly found in 
    the disk. If their positions are uncorrelated, the number of pairs should be proportional to some power of the intercluster 
    distance or separation. 
     
    The study of the distribution of the distance between two randomly chosen points within a sphere (or its two-dimensional 
    analogue, a circle) is a non-trivial problem and analytical formulae have repeatedly been found in the context of diverse 
    fields: Mathematics, Physics, Chemistry, Biology. The main advantage of using analytical techniques is that results are exact 
    and applicable as first order approximations to study complex real situations. The results for a disk (or a circle) may have 
    been known since the late 19th century (Crofton 1885) and they have been rediscovered in multiple occasions later (see e.g. 
    Garwood 1947; Hammersley 1950; Garwood and Tanner 1958; Barton et al. 1963; Solomon 1978; Gill et al. 2000). Regarding 
    the case of a sphere, it was first studied by Deltheil (1926) and later by Hammersley (1950) and Lord (1954). In the context 
    of Astrophysics, the pair distribution within a sphere was studied by Saiyan (1996) using Lord's results. The more general 
    problem of the distribution of distance between points independently and uniformly distributed in an arbitrary region has 
    been revisited by Alagar (1976). Here we review results for a circle and a sphere following Hammersley (1950). 

\begin{figure}[!htb]
      \centering
        \includegraphics[width=\columnwidth]{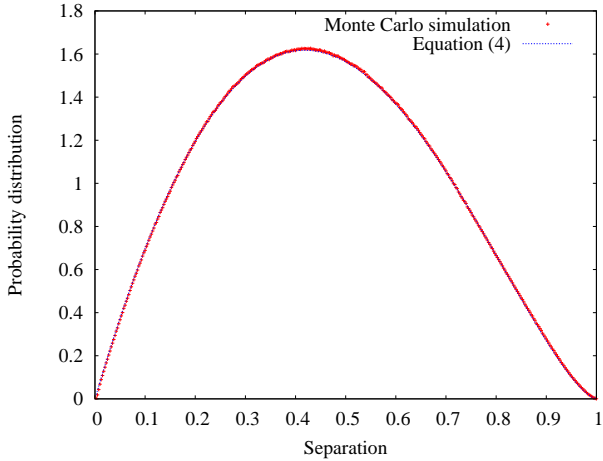}
        \caption{The distribution of distance in a unit-diameter disk. The solid curve shows Eq. (\ref{diskP}). The points 
                 are the result of the numerical experiment described in the text and provide the distribution of the distance 
                 between two independent random points each uniform inside a slice. The two data sets fully overlap.}  
      \label{disk}
    \end{figure}

    Let us consider a sample of random points on a unit-diameter disk (the inside of the unit-diameter circle), we want to find 
    the distribution of Euclidean distances or separations between two points at random on the disk. On a unit-diameter disk, the 
    probability that two points will be found at a separation $s$ is non-zero if $0 \leq s \leq 1$. The probability density 
    distribution or probability distribution of distances between points randomly distributed within a circle is given by the 
    so-called formula of Hammersley (1950):
    \begin{equation}
       P_{\rm disk}(s) = \frac{16 \ s}{\pi} \ (\arccos s - s \ \sqrt{(1 - s^2)})\,. \label{diskP}
    \end{equation}
    This function is displayed in Fig. \ref{disk} and it behaves almost linearly near $s = 0$ (small separations) and decays as 
    $(1 - s)^{3/2}$ near $s = 1$ (large separations). The most probable separation (maximum of the distribution) is approximately 
    $s = 0.418$ which is less than the radius of the disk. But, how reliable is this analytical expression? The correctness of 
    this theoretical result can be tested by generating a sample of random points (10$^4$ in our case) within a cylindrical slice 
    of radius $R_d$ and thickness $h$, with $h \ll R_d$, the Galactic disk. If we compute the distribution of the distance between 
    two random points picked within the slice using a Monte Carlo-type calculation (averaging the results of 10$^4$ experiments)
    and normalize, we obtain the points in Fig. \ref{disk}. Even if we allowed for non-zero thickness, the overlap is very good, 
    therefore and for separations $s \ll R_d$, the number of pairs grows linearly with $s$. For $h$ = 0 the agreement between the 
    Monte Carlo experiment and the results from Eq. (\ref{diskP}) is, in fact, perfect but for non-zero thickness and small 
    separations, significant deviations can be identified (see below).
%
%---------------------------------------------------------------------------------------------------------------------------------
%
%
%---------------------------------------------------------------------------------------------------------------------------------
%

\begin{figure}[!htb]
      \centering
        \includegraphics[width=\columnwidth]{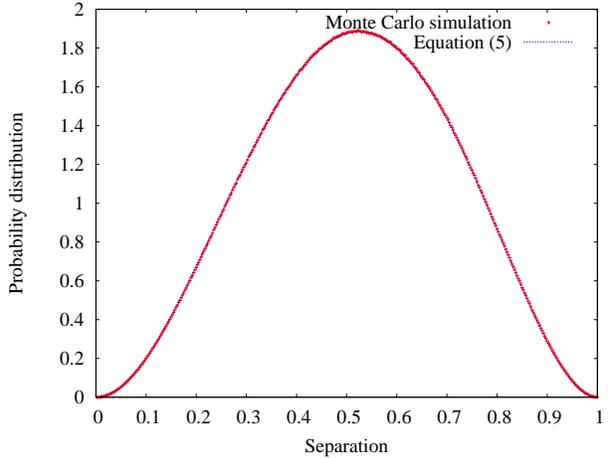}
        \caption{The distribution of distance in a unit-diameter ball. The solid curve shows Eq. (\ref{ballP}). The points 
                 are the result of the numerical experiment described in the text. As in Fig. \ref{disk}, the two data sets fully 
                 overlap.}
      \label{ball}
    \end{figure}

    Now, let us consider the inside of a unit-diameter sphere, what in mathematical terms is called the ball of unit diameter. 
    The probability density distribution for the distance between random points in a unit-diameter ball is zero unless $0 \leq s 
    \leq 1$. Again following Hammersley (1950), it is given by: 
    \begin{equation}
       P_{\rm ball}(s) = 12 \ s^2 \ (1 - s)^2 \ (2 + s)\,. \label{ballP}
    \end{equation}
    This is displayed in Fig. \ref{ball}. Equation (\ref{ballP}) grows quadratically near $s = 0$ and decays similarly near $s = 
    1$; the most probable separation, approximately $s = 0.525$, is larger than the radius of the ball. Figures \ref{disk} and 
    \ref{ball} show that whereas $P_{\rm disk}(s)$ is biased to the left, $P_{\rm ball}(s)$ is biased to the right. As in the 
    previous case, a Monte Carlo-type calculation (average of 10$^4$ experiments, each one including 10$^4$ points) shows that Eq. 
    (\ref{ballP}) is able to reproduce the distribution of the distance between two random points picked within a sphere of radius 
    $R_s$. For separations $s \ll R_s$, the number of pairs increases as the square of the separation. 
%
%---------------------------------------------------------------------------------------------------------------------------------
%
%
%---------------------------------------------------------------------------------------------------------------------------------
%

\begin{figure}[!htb]
      \centering
        \includegraphics[width=\columnwidth]{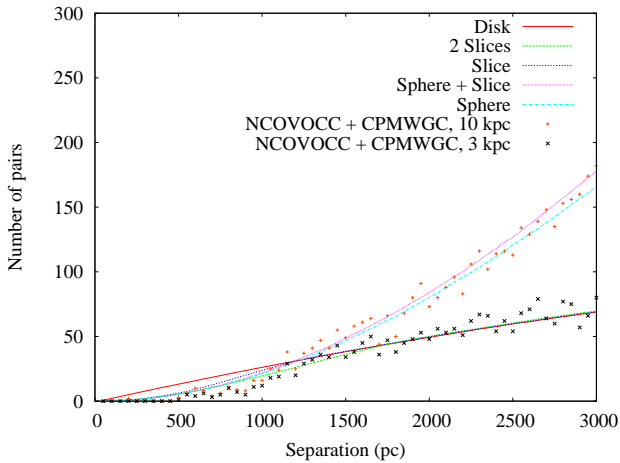}
        \caption{The distribution of distances between points from two independent samples randomly distributed within several 
                 volumes. The lines have been obtained using Monte Carlo-type calculations. Two independent random point 
                 distributions within a disk ($h$ = 0), continuous (red) line. Two independent random point distributions within 
                 two coplanar cylindrical slices ($h_A$ = 350 pc, $h_B$ = 1 kpc), {\small -- -- --} (green) line. Two independent 
                 random point distributions within a slice ($h$ = 1 kpc), {\small - - -} (dark blue) line. Two independent random 
                 point distributions, the first one within a slice ($h$ = 350 pc) and the second one within a sphere, \ldots 
                 (pink) line. Two independent random point distributions within one sphere, {\small - . -} (light blue) line. The 
                 radii of the disk, slices and spheres are equal to 10 kpc. For reference, we include real data for clusters found 
                 within 10 kpc from the Sun (+ signs) and 3 kpc from the Sun ($\times$ signs).}
      \label{2Dcases}
    \end{figure}

    If we consider two distinct, fully independent, completely unrelated point distributions, A and B, within a cylindrical slice 
    of radius $R_d$ and thickness $h$, with $h \ll R_d$, the distance probability density distribution for pairs of points, one 
    in A and another one in B, follows Eq. (\ref{diskP}). In the case of a sphere, it is described by Eq. (\ref{ballP}). 
    For a cylindrical slice and if the pair separation is significantly smaller than the thickness of the slice, the situation 
    locally resembles that of a sphere and the number of pairs increases as the square of the separation. In Fig. \ref{2Dcases} 
    we plot the results from several Monte Carlo-type calculations similar to the ones presented above and designed to further 
    explore this behaviour: the radius of the slice/sphere is 10 kpc and for the slice, the thickness takes values 0, 350 pc and 
    1 kpc. In the figure, the distribution of distances between points from two independent random point distributions within a 
    disk ($h$ = 0) appears as a (red) continuous line; two independent random point distributions within two coplanar cylindrical 
    slices ($h_A$ = 350 pc, $h_B$ = 1 kpc) are displayed as a {\small -- -- --} (green) line; two independent random point 
    distributions within a slice ($h$ = 1 kpc) appear as a {\small - - -} (dark blue) line; two independent random point 
    distributions, the first one within a slice ($h$ = 350 pc) and the second one within a sphere are displayed as a \ldots (pink) 
    line; two independent random point distributions within one sphere as a {\small - . -} (light blue) line. In advance of our 
    analysis of actual data presented in the following sections, we also include the observed (raw) distribution of distances for 
    real clusters (see below). The probability distribution of separations for pairs of objects (globular+open cluster) located 
    within 10 kpc from the Sun appears to follow the behaviour described by Eq. (\ref{ballP}) across the entire separation 
    spectrum in Fig. \ref{2Dcases}. If we restrict the sample to objects found within 3 kpc from the Sun, the spatial shape of the 
    sample resembles that of a slice and the probability distribution is better described by Eq. (\ref{diskP}) instead. However, 
    if $s < h$, the probability distribution appears to grow quadratically with the separation (see Fig. \ref{2Dcases}) although 
    some large fluctuations are observed for very small separations.
%
%---------------------------------------------------------------------------------------------------------------------------------
%
%
%---------------------------------------------------------------------------------------------------------------------------------
%

    The non-collisional scenario in which globular and open clusters never interact produces two regimes:
    \begin{itemize}
       \item The distribution of distances between globular and open clusters above a separation of $\sim$1250 pc follows the 
             linear behaviour associated to Eq. (\ref{diskP}) near $s = 0$.
       \item For smaller separations, the distribution of distances resembles the behaviour from Eq. (\ref{ballP}) near 
             $s = 0$.
    \end{itemize}
    This is the response observed for the model including two slices of different thickness, for example. The behaviour at small 
    separations may be regarded as unexpected but if $s \ll h$, locally speaking, the space surrounding the pair is less like a 
    disk and more like a sphere. That explains the quadratic evolution of the distribution of distances at small $s$. If the 
    non-collisional scenario is valid, the behaviour of the distribution of distances for $s <$ 1250 pc must be quadratic; 
    deviations from this theoretical behaviour must be the result of the mechanism pointed out in Sect. 2, the presence of young 
    open clusters formed by the passage of globular clusters and/or perhaps the association of some old open clusters to an 
    accreted globular cluster. In addition, completeness issues, uncertainties and biases plague observational data, and may 
    contribute deviations on their own. On the other hand, interactions between GMCs and star clusters may also have a role, 
    disrupting open clusters and truncating globular clusters. In the following, we will summarize the role of GMCs and try to 
    minimize all the adverse effects in order to obtain reasonably solid conclusions. 

 \section{Giant Molecular Cloud-star cluster interactions}
    Although somewhat unrelated to the issues discussed so far, it can be argued that GMCs not cluster-cluster interactions are 
    expected to play a dominant role on the dynamics and/or disruption of globular clusters and, by extension, open clusters 
    (e.g. Gieles et al. 2006). It is a well known result that a single close encounter or tidal shock between a GMC and an open 
    cluster can completely destroy the cluster (e.g. Spitzer 1958; Spitzer and Chevalier 1973; Surdin 1997). Newly born open 
    clusters with radii $>$ 2.5 pc are not expected to survive for long in the clumpy environment of their natal molecular clouds 
    (Long 1989). On the other hand, even if it is true that GMCs are very massive (10$^3$-10$^7 M_{\odot}$) and their overall 
    gravitational effects on star clusters are far from negligible (e.g. Terlevich 1987), their characteristic lifetimes are as 
    short as 27$\pm$12 Myr (Murray 2011) which compare unfavourably with the typical lifetimes of both open and globular clusters 
    (0.1-100 Gyr). Statistically speaking, encounters between open and globular clusters are much more probable than those 
    between GMCs and any type of star cluster. The very short lifetimes of GMCs make them somewhat inefficient as long-term 
    perturbers of star clusters but they are quite short-term effective at the cluster birth site (Elmegreen and Hunter 2010). 
    Therefore, GMCs can be regarded as very effective in disrupting very young open clusters but their effectiveness must be 
    significantly lower for older open clusters. In any case, globular-open cluster interactions and GMC interactions with star 
    clusters can both take place and play a dynamical role although, in strict terms of time-scale, globular-open cluster 
    interactions are expected to be more frequent.

    As a quantitative example, the Bell Laboratories $^{13}$CO Survey has studied the first quadrant of the Milky Way to identify 
    GMCs (Lee et al. 2001). The associated catalogue includes 1400 clouds with a virial mass range of 10$^2$-10$^7 M_{\odot}$ and 
    a likely size range of 5-200 pc; 56 objects have masses $> 10^5 M_{\odot}$ which host over 85~\% of the total mass and 
    most GMCs have masses in the range 10$^3$-10$^4 M_{\odot}$ (see fig. 3, Stark and Lee 2006). Data from this survey show that 
    the scale height of the most massive GMCs is below that of smaller clouds which is about 35 pc (Stark and Lee 2005); i.e., the 
    scale height declines with increasing cloud mass. Also, most massive GMCs are more concentrated toward spiral arms than 
    smaller clouds (Stark and Lee 2006). These numbers suggest that the current population of GMCs of all sizes in the Milky Way 
    is nearly 6000 but only about 250 objects are truly massive. This number is comparable to that of the globular clusters. The 
    probability of a close encounter between a massive GMC and a globular cluster is rather small but not negligible because the 
    size of large GMCs is nearly 200 pc. Such an encounter may trigger star-formation within the GMC. At any given time in the 
    history of the Galaxy, the number of smaller GMCs is clearly lower than the total number of open clusters and their masses, 
    for the most typical, small GMCs, are similar to those of the largest open clusters. However, their lifetimes are shorter 
    than the typical evaporation time of open clusters; i.e. $<$ 100 Myr. These facts strongly suggest that GMC-cluster 
    interactions are not overly dominant with respect to globular-open cluster interactions; in fact, both processes appear to be 
    cooperating in the sense that they equally contribute to the tidal evolution of star clusters. On the other hand, the number 
    of GMCs is insufficient to heat the Galactic disk appropriately (H\"anninen and Flynn 2002). Most GMCs are located at 
    Galactocentric distances in the range 4-7 kpc (Anderson et al. 2012). In any case, most GMCs would destroy themselves via star 
    formation before they encounter a cluster, either globular or open; also, the vast majority of the gas is in the form of large 
    clouds which indicates that cloud-cloud interactions are completely negligible, probably because of their short lifetimes. If 
    we apply Eq. (\ref{timescale1}) to compute the characteristic time-scale for GMC-open cluster interactions assuming that the 
    number of GMCs of all sizes within 1 kpc from the Sun is 50 (it must be close to half the number of young open clusters within 
    the same volume which is 92, see below, because a single cloud may produce several clusters), the velocity ratio must be close 
    to 1 as open clusters are born out of GMCs, $r$ = 1 kpc, $h$ = 35 pc and $r_t$ = 35 pc, we obtain nearly 0.2 Myr which is 
    similar to the value found in Sect. 2 for globular-open cluster interactions. In the case of globular clusters, the time-scale 
    is longer because the ratio of velocities is, in general, $> 1$, the other variables taking the same values. The topic of 
    encounters between molecular clouds and globular clusters has been studied by Surdin (1997). He found that globular clusters 
    moving in retrograde orbits are virtually unaffected by the disruptive action of GMCs; as for prograde orbits, a single 
    encounter cannot destroy the cluster but multiple encounters can. Surdin (1997) ignored the effects of the globular cluster on 
    the GMC.

\begin{figure}[!htb]
      \centering
        \includegraphics[width=\columnwidth]{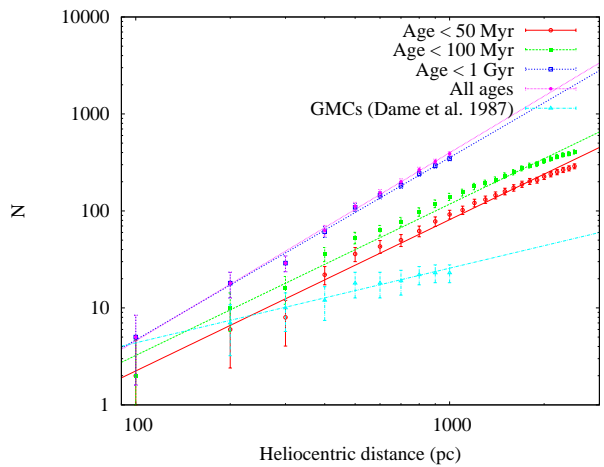}
        \caption{Number of open clusters, $N$, in NCOVOCC as a function of their heliocentric distance, $d$, for various age 
                 groups. The associated best least squares fittings are also plotted. The power-law index evolves over time and 
                 for embedded clusters it must be close to that found for GMCs, 0.77$\pm$0.06. The surface density of open 
                 clusters eventually becomes uniform but starts growing as $\sim~d^{-1}$ for very young open clusters. Error bars 
                 display the 1-$\sigma$ Poisson error (see Sect. 5.2 for details).}  
      \label{com}
    \end{figure}

    As for the local GMC sample, the classical study of Dame et al. (1987) shows that the distribution of GMCs is rather 
    irregular. The molecular mass within 1 kpc from the Sun is four times greater in the first and second quadrants than in the
    third and fourth (see table 2 and fig. 7 in Dame et al. 1987). Nearly all the clouds within 1 kpc in the first and fourth
    quadrants apparently lie on a fairly straight ridge likely associated to the local spiral arm (see fig. 7 in Dame et al. 
    1987). This section of the local spiral arm is probably a branch of the Perseus arm (see fig 10 in Xu et al. 2013). The 
    sample in Dame et al. (1987) includes about 20 major clouds with masses in the range 0.03-8.7 $\times 10^5 M_{\odot}$. This 
    sample is likely complete for that range of masses. The surface density of GMCs in the Solar Neighbourhood is far from 
    uniform. The number of GMCs as a function of the heliocentric distance follows a power-law with an index of 0.77$\pm$0.06 and 
    a correlation coefficient of 0.981 (see Fig. \ref{com}). This non-uniform distribution of the GMCs has strong effects on the 
    surface density of newly born open clusters (age $<$ 100 Myr) as open clusters are formed out of GMCs.
%
%---------------------------------------------------------------------------------------------------------------------------------
%
%
%---------------------------------------------------------------------------------------------------------------------------------
%

 \section{Data and actual distribution of separations}
    How do the theoretical predictions from both the realistic collisional (Sect. 2) and the ideal non-collisional (Sect. 3) 
    scenarios compare with actual observational data? Here, we present the samples used in our analysis and compute the current 
    intercluster distance spread; we also discuss some relevant globular clusters.

    \subsection{The samples}
       A first step in testing the predictions posed above is to carefully examine the intercluster distance spread using real 
       data. In order to carry out a systematic study of the globular-open cluster separation distribution we use three on-line 
       public databases:

       {\it NCOVOCC.} The {\it New Catalogue of Optically Visible Open Clusters and Candidates}\footnote{http://www.astro.iag.usp.br/$\sim$wilton/} 
       (NCOVOCC, Dias et al. 2002) is widely used in open cluster studies. The January 2013 version (v3.3, Dias 2013) of NCOVOCC 
       includes 2174 open clusters and 1629 of these (74.9~\%) have known distances and ages. This is the subsample used in 
       our analysis. There are 389 open clusters of all ages within 1 kpc from the Sun in NCOVOCC. In addition, we have used data 
       from the {\it Open Cluster Database}\footnote{http://www.univie.ac.at/webda/} (WEBDA, Mermilliod and Paunzen 2003), another
       widely used open cluster database, to check for consistency with NCOVOCC but we have not included any results from this 
       database because the currently available version (May 2013, Paunzen and Mermilliod 2013) is less extensive than NCOVOCC as 
       it includes 1755 objects with 1059 (60.3~\%) of them with both known age and known distance. 

       {\it CPMWGC.} The {\it Catalog of Parameters for Milky Way Globular Clusters}\footnote{http://www.physics.mcmaster.ca/Globular.html} 
       (CPMWGC, Harris 1996) plays a role similar to that of NCOVOCC or WEBDA for open clusters but in the field of Galactic 
       globular clusters. The latest revision of this catalogue (December 2010, Harris 2010) contains basic information on 
       distances, velocities, metallicities, colours and dynamical parameters for 157 objects classified as globular clusters in 
       the Milky Way. One of these objects, GLIMPSE-C01, has been recently found to be an intermediate-age (400-800 Myr) open 
       cluster (Davies et al. 2011) not a globular cluster and, for this reason, it has been removed from our calculations. 
       Therefore, 156 objects are included in this sample. There are currently 57 globular clusters within 1 kpc from the 
       Galactic disk (36.5~\%), i.e. within the thick disk. No globular clusters have been found within 1 kpc from the Sun. 
       CPMWGC does not include the controversial globular cluster candidate FSR 1767 (see Froebrich et al. 2009 and 
       Bonatto et al. 2009 for details) as Harris considers that the currently available evidence on this object is not yet 
       convincing. FSR 1767 will be the subject of a more detailed analysis in Sect. 8.   

       {\it GGCD+.} The {\it Galactic Globular Cluster Database}\footnote{http://gclusters.altervista.org} (GGCD, Castellani 
       2008) is based on Harris (2010). The latest version (December 2011, Castellani 2011) includes 157 objects. This dataset 
       has been enhanced with additional data collected from the literature for objects discovered within the last few years and 
       GLIMPSE-C01 has been deleted from the list (see above). The total number of objects in this enhanced version is 186 and it 
       considers both globulars and close satellite dwarf galaxies. The disputed globular cluster candidate FSR 1767 (see Sect. 
       8 for details) is part of GGCD. Therefore, GGCD+ includes GGCD and 30 additional objects. With the remarkable exception of 
       FSR 1767, the impact of GGCD+ on our conclusions is negligible as the vast majority of the newly discovered objects are 
       located far from both the Sun and the Milky Way plane.

       We can safely assume that the list of known galactic globular clusters is complete at the 90~\% level (e.g. Harris 
       1991) or even the 94~\% level (Ivanov et al. 2005) but this is however not the case, in general, for 
       open clusters. The samples considered above are not volume-limited and it can be easily argued that they are strongly 
       biased in favor of young objects as older and dimmer open clusters are more difficult to identify. The study of 
       completeness in general open cluster samples customarily assumes that the surface density of open clusters in the Solar 
       Neighbourhood is uniform (Battinelli and Capuzzo-Dolcetta 1989, 1991). This hypothesis implies $N \propto d^2$, where $N$ 
       is the number of clusters and $d$ is a given heliocentric distance. In their papers, it was found that for open clusters 
       within 2 kpc from the Sun and brighter than $M_V$ = -4.5 (see fig. 5 in Battinelli et al. 1994), 
       the assumption of uniform average number density of open clusters was matched well by the observational results. More 
       recently, Piskunov et al. (2006) have concluded that, assuming uniform density, the completeness limit for clusters of any 
       age could be 0.85 kpc. De la Fuente Marcos and de la Fuente Marcos (2009) found similar results: for objects in the age 
       range 1-1000 Myr located within 0.9 pc from the Sun, the power-law index is 1.96$\pm$0.05 with a correlation coefficient 
       of 0.998. For our current sample of open clusters from NCOVOCC, a similar analysis (see Fig. \ref{com}) gives 
       1.94$\pm$0.05 with a correlation coefficient of 0.997 for open clusters of all ages located within 1 kpc from the Sun. 
       Therefore, our volume-limited open cluster samples are likely to be at least 90~\% complete even for older clusters 
       if the radial distance is restricted to about 1 kpc; unfortunately and as pointed out above, there are no globular 
       clusters within 1 kpc from the Sun. Based on this completeness limit, the actual number of open clusters of all ages 
       within 1 kpc from the Sun could be as high as $\sim$400. Given the fact that there are 389 open clusters of all ages 
       within 1 kpc from the Sun in NCOVOCC, it can be claimed that this sample is nearly 100~\% complete. As for younger 
       objects, which are most relevant to our present research in order to discuss the role of globular clusters in open cluster 
       formation, de la Fuente Marcos and de la Fuente Marcos (2008a, 2009) have found reasonable evidence that the age 
       distribution of young open clusters ($\leq$ 100 Myr) within 2.5 kpc from the Sun is not severely affected by detection 
       limits and it is likely nearly 90~\% complete too. This is mainly due to the fact that young star clusters contain 
       very luminous stars that can be seen at large distances and the effect of incompleteness is, consistently, smaller for 
       them. For our present sample (see Fig. \ref{com}), the situation is similar although the uniform surface density model 
       cannot be invoked in this case as the power-law index is now intermediate (1.56$\pm$0.04 with a correlation coefficient of 
       0.992) between that of the GMCs (see Sect. 4) and that of the general, non-age-limited open cluster sample. The 
       evolutionary behaviour of the value of the power-law index reflects the tendency for an increase in the average distance 
       between surviving open clusters over time.

       Although the quality of the data in the above databases is rather inhomogeneous and the data compilation is necessarily 
       incomplete as it is the research in the field of star clusters in the Milky Way, we consider the samples used in this work 
       as sufficiently complete and reliable, or at least as the best available for the purpose of this research. The current 
       status of the accuracy of open cluster parameters has been reviewed by Paunzen and Netopil (2006). They found that 
       distances are rather well determined: the absolute error is less than 20~\% for nearly 80~\% of the best 
       studied open clusters. The situation is quite the opposite for ages as only 11~\% of the investigated open clusters 
       have errors less than 20~\% and 30~\% exhibit an absolute error in the estimate of the age larger than 50~\%. 
       In addition to that, Paunzen and Netopil (2006) show that the errors are age dependent and the ones associated to 
       young clusters are sometimes significantly larger (in average) than those of older objects, with 20--30~\% in the 
       best possible cases. In a large sample, these errors are non-homogeneous because different methods have been used by 
       different authors to calculate the ages. The situation is likely similar for globular clusters. 
 
    \subsection{Cluster separation analysis}
\begin{figure*}[!htb]
         \centering
           \includegraphics[width=\textwidth]{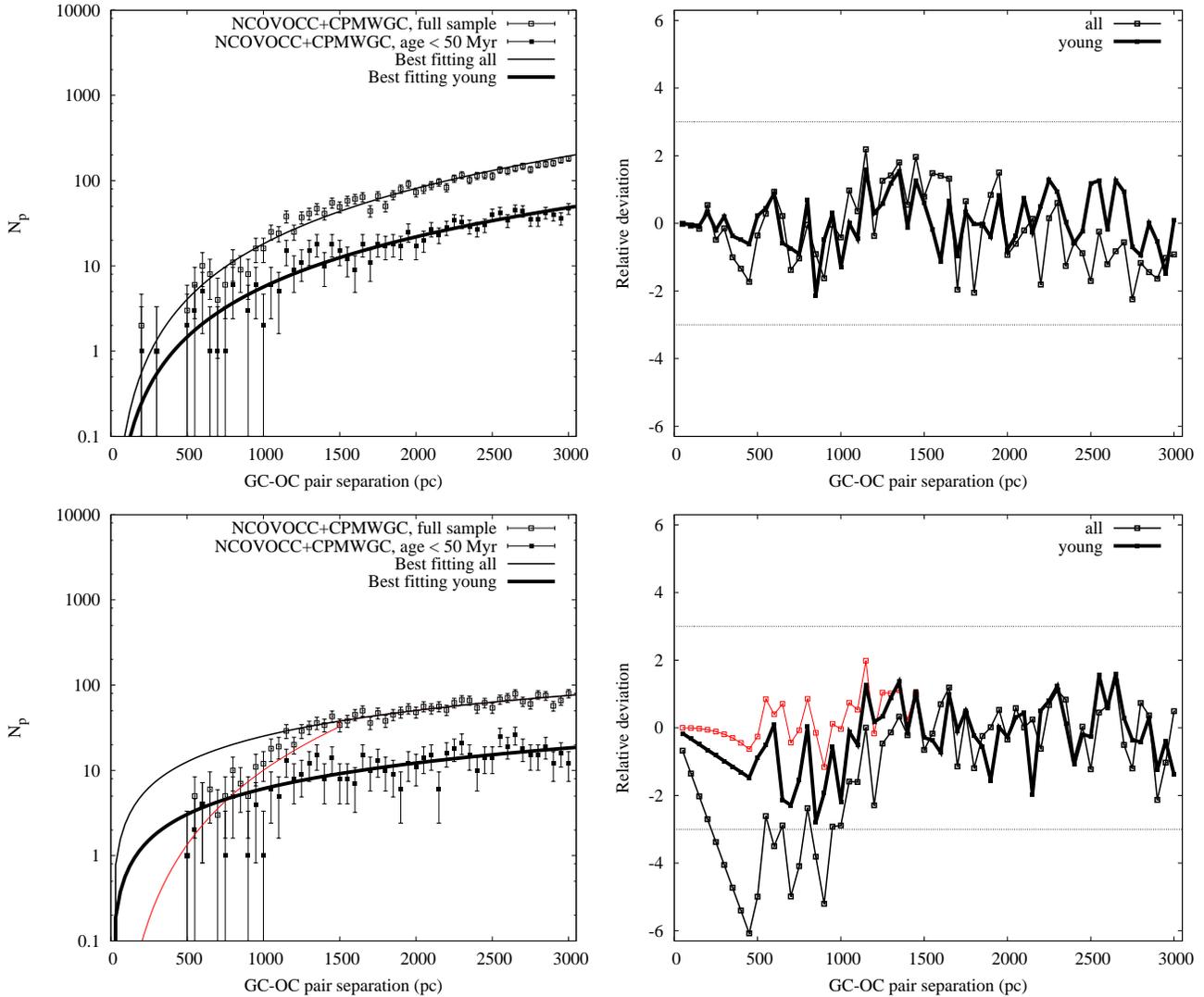}
           \caption{Number of globular cluster-open cluster pairs as a function of their separation for NCOVOCC+CPMWGC. The 
                    intercluster distance distribution is shown here for all the open clusters (empty squares, thin lines) and 
                    just the young ones (filled squares, thick lines, age $<$ 50 Myr). The actual distributions of separations 
                    are displayed on the left-hand panels, deviations from the non-collisional theoretical models (thin, thick 
                    lines, Sect. 3) in units of the standard deviation, $\sigma$, appear on the right-hand ones. The top panels 
                    are for objects within 10 kpc from the Sun, at the bottom we display data for objects within 3 kpc from the 
                    Sun. Error bars display the 1$\sigma$ Poisson error (see the text for details). For objects within 3 kpc from 
                    the Sun and separations closer than $\sim$1 kpc obvious deviations from the theoretical expectations are 
                    observed. There are no close pairs ($s < 450$ pc) within 3 kpc from the Sun. The red data correspond to the 
                    $\sim s^{3}$ model discussed in the text. The 3$\sigma$ limits are also indicated for the right-hand panels.}
         \label{evidence}
       \end{figure*}

       Figure \ref{evidence} shows the distribution of distances for the NCOVOCC+CPMWGC databases introduced above: the number of 
       globular cluster-open cluster pairs ($N_p$) as a function of their separation. This separation has been calculated using 
       the Euclidean distance between two points in three-dimensional space with Galactic Heliocentric Cartesian coordinates. 
       Analytical models based on the results presented in Sect. 3 are included for comparison. We assume Poissonian number 
       counts and error bars display the 1$\sigma$ Poisson error, $\sqrt{N_p}$. However, in some cases we deal with small $N_p$; 
       the inadequacy of the $\sqrt{N_p}$ approximation for small number of counts has been known for long (e.g. Regener 1951). 
       Here we use the approximation given by Gehrels (1986) when $N_p < 21$: $\sigma \sim 1 + \sqrt{0.75 + N_p}$. The top panels 
       in Fig. \ref{evidence} show the results for objects within 10 kpc from the Sun. If the two random samples are enclosed by 
       a sphere, then $N_p \propto s^2$ ($s \ll$ sphere radius, see Sect. 3 for details) but in a disk setup and if the volume 
       considered is much smaller than the one associated to the thickness of the disk, it also resembles a spherical setup and 
       the $N_p \propto s^2$ behaviour is expected (see Sect. 3 for the slice model). The figure (left-hand panel) exhibits the 
       expected quadratic trend for both young open clusters (age $<$ 50 Myr, thick line) and for the general sample (without age 
       restriction, thin line). The full sample gives a power-law index of 2.16$\pm$0.07 with a correlation coefficient of 0.977; 
       for young open clusters we obtain a power-law index of 1.95$\pm$0.13 with a correlation coefficient of 0.919. The 
       deviations from the power-law model in units of $\sigma$, the standard deviation, are shown in the right-hand panel. 
       Theoretical predictions from the non-collisional model and observational data agree within the error limits; the deviations 
       are only marginally significant. However, it must be pointed out that open cluster samples within 10 kpc from the Sun are 
       very incomplete for any age range; the good agreement could easily be the result of heavy incompleteness. If we consider 
       the pair subset characterized by intercluster distances $<$ 3 kpc (bottom panels), we are still well above the distinctive 
       scale heights of both thick (1 kpc) and thin (350 pc) disks (e.g. Sparke and Gallagher 2007). We must however point out 
       that it has recently been argued that the Milky Way has no distinct thick disk but a continuous and monotonic distribution 
       of disk thicknesses (Bovy et al. 2012). For two fully uncorrelated (random) samples of data points contained within a disk, 
       the number of pairs separated by a given distance $s$ ($\ll$ disk radius) grows directly proportional to their separation, 
       $N_p \propto s$ but only if the separation is larger than the characteristic thickness of the distribution (see the 
       discussion in Sect. 3 on the two regimes). In principle and for small separations, if globular and open clusters are fully 
       unrelated we should observe the linear behaviour as only globular clusters located within the disk contribute in this 
       regime; however, for very small separations (smaller than the thickness of the disk) the scenario locally resembles that of 
       a spherical configuration and the $N_p \propto s^2$ should be expected. Figure \ref{evidence} (bottom panels) is consistent 
       with linear growth (correlation coefficients of 0.994 and 0.948 for the general and young samples, respectively) for 
       separations larger than 1 kpc but appears to be incompatible with that model for smaller separations and, more importantly, 
       no pairs closer than 450 pc are observed. If we consider a $N_p \propto s^2$ model in the range 0-1 kpc, it fails to 
       account for the irregular behaviour displayed and the lack of close pairs. The best fitting to the data for small 
       separations is obtained for a power-law index of 2.9$\pm$0.3 with a correlation coefficient of 0.927. This corresponds to 
       the red data displayed in the bottom panels (Fig. \ref{evidence}). 

       Small deviations (specially those associated to small numbers of pairs) can be regarded as statistical fluctuations and 
       they may not be significant; however, the overall lack of pairs observed at small separations appears to be a real feature 
       not attributable to fluctuations. This is one of the predictions of the mechanism described in Sect. 2. This scarcity of 
       close pairs can be interpreted as the result of hypothetical globular-open cluster dynamical interactions. These 
       interactions may be destructive (cluster harassment), translating into reduced numbers of pairs at small separations, or 
       constructive, as in the globular clusters passing through the disk and triggering formation of open clusters scenario. The
       results from Fig. \ref{evidence} are sensitive only to destructive interactions and they strongly suggest that globular
       clusters are not responsible for the destruction of a sizable fraction of young open clusters. The distribution of 
       distances for pairs including young open clusters appears to be compatible with predictions from the non-collisional 
       model. Unfortunately, it is also true that the quality of the available observational data may be behind the observed 
       features and they could be artefacts. In Sect. 5.1 we emphasized the fact that the non-age-limited open cluster sample 
       is rather complete (90~\% or better) for heliocentric distances $<$ 1 kpc. We also pointed out that for young 
       clusters our sample could be reasonably complete (perhaps up to 90~\%) even if the heliocentric distance is in the 
       range 2.5-3.0 kpc. The distribution of young open clusters is far from uniform as it is shaped by that of their parent 
       GMCs; in contrast, the distribution of open clusters without age restriction appears to be uniform. The detectability of 
       older clusters decreases significantly with the distance. Our general open cluster sample with heliocentric distances $<$ 
       3 kpc includes 1274 objects but our uniform surface density model predicts 3600 open clusters; therefore, our sample is 
       35.4~\% complete. The vast majority of the missing objects must be old open clusters, which are uniformly 
       distributed. The impact of the missing clusters on the distribution of distances studied above must be evenly scattered; 
       i.e., all the bins in the histogram must be equally affected by the lack of these uniformly distributed open clusters not 
       a specific range of separations. Our previous analysis indicates that statistically significant deviations from the 
       non-collisional scenario in the range 3-6$\sigma$ are observed for cluster pairs with heliocentric distances $<$ 3 kpc. 
       These deviations affect the range of separations 200-1000 pc. The statistically significant deviations observed cannot be 
       just the result of incompleteness. The sample with $d < 10$ kpc is far more incomplete than that of $d < 3$ kpc for both
       young and older clusters and its results are still compatible with the non-collisional scenario within the error limits. 

       In order to find out the cause of this intriguing behaviour, let us further study the number of open clusters found within 
       1 kpc of any globular cluster (results are summarized in Table \ref{fraction}).
%
%---------------------------------------------------------------------------------------------------------------------------------
%
       \begin{table}[!htb]
         \fontsize{8}{10pt}\selectfont
         \tabcolsep 0.2truecm  
         \begin{center}
           \caption{Fraction of globular clusters with one or more companions within 1 kpc
                    from the various datasets and subsamples.}
           \resizebox{\linewidth}{0.19\linewidth}{ 
           \begin{tabular}{|c|c|cccc|}
             \hline  
                            & Dataset        &      Non       & $d < 10$ kpc & $d < 3$ kpc \\
                            &                & volume-limited &              &             \\
             \hline 
                Full        & NCOVOCC+CPMWGC &  8.3 \%        & 15.1 \%      & 100 \%      \\
              datasets      & NCOVOCC+GGCD+  & 10.2 \%        & 22.4 \%      & 100 \%      \\
             \hline 
                  Young     & NCOVOCC+CPMWGC &  5.8 \%        & 10.5 \%      & 100 \%      \\
              open clusters & NCOVOCC+GGCD+  &  7.5 \%        & 16.5 \%      & 100 \%      \\
             \hline 
                   Old      & NCOVOCC+CPMWGC &  4.5 \%        &  8.1 \%      & 66.7 \%     \\
              open clusters & NCOVOCC+GGCD+  &  5.9 \%        & 12.9 \%      & 75 \%       \\
             \hline
           \end{tabular}
           }
           \label{fraction}
         \end{center}
       \end{table}
%
%---------------------------------------------------------------------------------------------------------------------------------
%
%
%---------------------------------------------------------------------------------------------------------------------------------
%
%
%---------------------------------------------------------------------------------------------------------------------------------
%

       {\bf Full datasets.}
       Using both NCOVOCC and CPMWGC we calculate the physical distance (not projected) between each globular cluster in CPMWGC 
       and every open cluster in the NCOVOCC sample, 254\,124 pairs. Here we only focus on the pair subset characterized by 
       intercluster distances $<$ 1 kpc that includes just 100 pairs (0.04~\%). Only 13 (8.3~\%) globular clusters out of 156 have 
       (one or more) open clusters within 1 kpc of their centres. Out of this subset, only three pairs (3~\%) are separated by 
       less than 300 pc. These rather close pairs are associated to the globular cluster 2MASS-GC01 (163 pc to van den Bergh 113, 
       174 pc to NGC 6603, 277 pc to NGC 6561). 2MASS-GC01 is located 3.6 kpc from the Sun. If we restrict our calculations to 
       globular clusters located within 10 kpc from the Sun (86 objects) we still obtain the same 100 pairs; the fraction of 
       globular clusters with relatively close ($<$ 1 kpc) open cluster companions goes up to 15.1~\%. However, if we further 
       restrict our calculations to the subsample with heliocentric distances $<$ 5 kpc (14 objects), the number of pairs goes 
       down to 95 and the fraction of globulars with open cluster companions is now 57~\%. Given the fact that globular cluster 
       samples within 10 kpc from the Sun are very likely 99~\% complete (see above), we may be tempted to conclude that it is 
       rather unusual for an open cluster to be found within 1 kpc of a globular cluster. However, general open cluster samples 
       are only reasonably complete within 1 kpc from the Sun (see above) and the fraction of globular clusters found within 3 kpc 
       from the Sun with open cluster companions closer than 1 kpc is 100~\% (three clusters with 75 companions). In the Solar 
       Neighbourhood, at least, it is normal to find open clusters relatively close to globular clusters.

       If we now perform a similar analysis for NCOVOCC with 1629 objects and GGCD+ with 186 objects, we get 302\,994 pairs in 
       total; out of them, 14 pairs are separated by less than 300 pc. The smallest separation, 36 pc, is found for the cluster 
       pair FSR 1767/Ruprecht 127. Following Innanen et al. (1972), for two clusters separated by a distance larger than three 
       times the outer radius of each cluster, the amount of mutual disruption is rather negligible. This is the case for this 
       pair, if real. Ten other open clusters are within 300 pc of FSR 1767. The non-volume-limited sample gives 307 pairs with 
       intercluster distances $<$ 1 kpc (0.10~\%); only 19 objects out of 186 (10.2~\%) have an open cluster within 1 
       kpc from their centres. FSR 1767 has 180 open clusters within 1 kpc from its centre. For the volume-limited sample 
       with radius 10 kpc, we still obtain 307 pairs with 19 objects out of 85 (22.4~\%) having an open cluster companion 
       within 1 kpc. That fraction raises to 63.2~\% for the volume-limited sample with radius 5 kpc in which 12 objects 
       out of 19 contribute 299 pairs. As in previous cases, the fraction of globular clusters within 3 kpc from the Sun with 
       open cluster companions closer than 1 kpc is 100~\% (four clusters with 264 companions). Results from this data set 
       are, in principle, compatible with those found above; the only difference is the presence of a clear outlier, FSR 1767.  

       {\bf Young open clusters.}
       For open clusters younger than 50 Myr and using NCOVOCC and CPMWGC, we obtain 32 pairs with nine globulars out of 86 
       (10.5~\%) having a close open cluster companion for globular clusters within 10 kpc from the Sun. The sample located 
       within 5 kpc from the Sun gives 30 pairs with seven globulars out of 14 (50~\%) having close companions. The 
       fraction of globular clusters within 3 kpc from the Sun with young open cluster companions closer than 1 kpc is 100~\% 
       (three clusters with 22 companions).

       For the dataset including data from NCOVOCC and GGCD+ and for globulars within 10 kpc from the Sun, we obtain 95 pairs 
       with 14 globulars out of 85 (16.5~\%) having a young open cluster companion within 1 kpc. For the sample located 
       within 5 kpc from the Sun we find 92 pairs with 11 globulars out of 19 having a young companion (57.9~\%). As usual, 
       the fraction of globular clusters within 3 kpc from the Sun with open cluster companions closer than 1 kpc is 100~\% 
       (4 clusters with 77 companions).

       {\bf Old open clusters}
       Restricting our analysis to open clusters older than 1 Gyr, the dataset including objects from NCOVOCC and CPMWGC gives 11 
       pairs with seven globulars out of 86 (8.1~\%) having a close old open cluster companion for objects within 10 kpc 
       from the Sun. The sample located within 5 kpc from the Sun still includes nine pairs with five globulars out of 14 
       (35.7~\%) having close companions. The fraction of globular clusters within 3 kpc from the Sun with old open cluster 
       companions closer than 1 kpc is 66.7~\% (two out of three clusters with six companions).

       For the dataset including data from NCOVOCC and GGCD+ and for globulars within 10 kpc from the Sun, we obtain 40 pairs 
       with 11 globulars out of 85 (12.9~\%) having an old open cluster companion within 1 kpc. For the sample located 
       within 5 kpc from the Sun we find 37 pairs with nine globulars out of 19 having an old companion (47.4~\%). The 
       fraction of globular clusters within 3 kpc from the Sun with old open cluster companions closer than 1 kpc is 75~\% 
       (three out of four clusters with 30 companions). 

    \subsection{Pair separation trends}
\begin{figure}[!htb]
         \centering
           \includegraphics[width=\columnwidth]{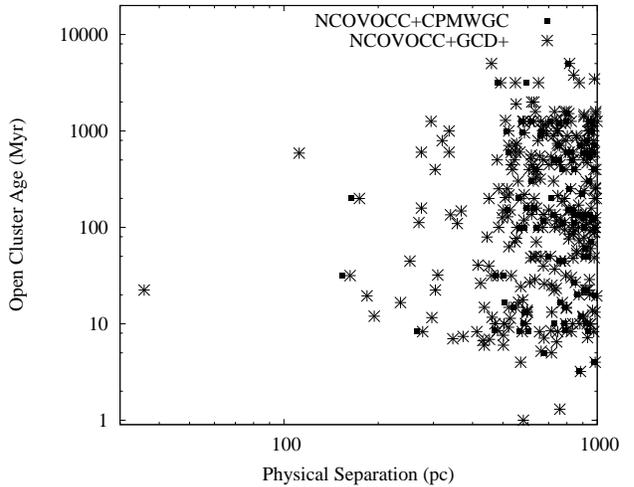}
           \caption{Age of the open cluster in the globular-open cluster pair as a function of its physical separation for the 
                    two datasets. In order to improve clarity, points from NCOVOCC+CPMWGC are shifted by 10 units in physical 
                    separation. See Table \ref{10pairs} for details on the closest pairs.}
         \label{pairs}
       \end{figure}

       If we represent the age of the open cluster in the pair as a function of the physical separation of the pair we observe 
       some obvious trends (see Fig. \ref{pairs}). With the exception of FSR 1767-Ruprecht 127, no pairs closer than 100 pc are 
       observed. As pointed out above, this is consistent with theoretical expectations if globular and open clusters interact. 
       The average tidal radius for globular clusters in our sample is 36.3 pc with a standard deviation of 27.5 pc. Therefore, 
       the characteristic cutoff distance to avoid significant disruption is nearly 110 pc. Open clusters located closer than the 
       cutoff distance are expected to be tidally disrupted by the passing globular cluster in a short time-scale. In general, it 
       is rather difficult to find any old open clusters relatively close to globular clusters. This may indicate that no old open 
       clusters are currently associated to globular clusters but the actual evidence is blurred by the fact that samples of old 
       open clusters are rather incomplete and also because old open clusters are just survivors and necessarily scarce. 
 
    \subsection{Some relevant globular clusters} 
       Here we summarize data on the best candidate clusters in Table \ref{candidates}. All of them are located towards the 
       Galactic centre and they are the closest globular clusters to the Sun. If not explicit, the reference for the data in this 
       section is Harris (2010).
     
       {\bf NGC 6121 (M 4).}
       Located in Scorpius, this nearby, 2.2 kpc, metal poor ([Fe/H] = -1.16) globular cluster shows evidence for two stellar 
       populations (Marino et al. 2008; Villanova and Geisler 2011). Its core radius is 1.2 pc, the tidal radius 33.2 pc and its 
       ellipticity is zero. The physical distance to this object has been recently revised and found to be 1.80$\pm$0.05 kpc 
       (Hendricks et al. 2012). Dinescu et al. (1999) found that for NGC 6121 the total velocity relative to the 
       Galactic centre is 201 km~s$^{-1}$ with a vertical velocity of -8 km~s$^{-1}$. No old open clusters are found close to 
       this object and the dynamical evidence for a tidally induced star formation episode associated with this particular object 
       is not obvious. The closest open cluster is ASCC 88 (15 Myr old) at 550 pc but there are 11 other clusters younger than
       100 Myr and with separations from NGC 6121 in the range 550-1000 pc. 

       {\bf NGC 6397.}
       Located in Ara, this nearby, 2.3 kpc, old (12 Gyr), metal poor ([Fe/H] = -2.02) globular cluster is undergoing core 
       collapse. Its core radius is 0.05 pc with a tidal radius of 10.6 pc and an ellipticity of 0.07. Its radial velocity is 
       18.8 km/s. A super-Li rich turn-off star has recently been found in this cluster (Koch et al. 2011) and it hosts two 
       stellar populations as its main sequence splits into two components (Milone et al. 2012). Kalirai et al. (2007) found 
       that for NGC 6397 the total velocity relative to the Galactic centre is 195 km~s$^{-1}$ with a vertical velocity of -140 
       km~s$^{-1}$; this cluster has experienced a recent disk-crossing shock during its passage through the Galactic plane 3.7 
       Myr ago (Chen and Chen 2010). Currently, there are nine young open clusters with separations from NGC 6397 in the range 
       500-1000 pc. NGC 6253, an old (5 Gyr), metal rich ([Fe/H] = +0.43) open cluster is found about 800 pc from NGC 6397 but 
       its radial velocity is very different, -29.4 km/s (Anthony-Twarog et al. 2010), and this makes any association between the 
       two star clusters highly unlikely.  

       {\bf NGC 6544.}
       Located in Sagittarius, NGC 6544 is one of the closest (3.0 kpc) and most compact globular clusters known. Its ellipticity 
       is the fourth highest among Galactic globulars, $e = 0.22$. With a core radius of 0.05 pc and a tidal radius of 1.9 pc, 
       its very concentrated structure could be the signature of a previous encounter with a giant molecular cloud. Its radial 
       velocity is -27$\pm$4 km/s and one of its neighbouring young open clusters may share it (separations and ages are shown in 
       parentheses): ASCC 93 (236 pc, -23.33 km/s, 16 Myr). Turner 4 (463 pc, 10 Myr), Ruprecht 140 (500 pc, 32 Myr), Pismis 24 
       (880 pc, 10 Myr), Turner 3 (942 pc, 20 Myr), Dias 5 (966 pc, 14 Myr) and Trumpler 33 (978 pc, 48 Myr) may also be related 
       to NGC 6544. 

       {\bf NGC 6656 (M 22).}
       Also found in Sagittarius, NGC 6656 is located 3.2 kpc from the Sun and has an ellipticity of 0.14. It has an unusually 
       large core (radius of 1.32 pc) that has been explained as the result of interactions between cluster stars and black holes
       (Strader et al. 2012). Its tidal radius is 29.7 pc and four open clusters younger than 100 Myr have separations from NGC 
       6656 in the range 500-1000 pc. 

       {\bf 2MASS-GC01/vdBergh 113/NGC 6561/NGC 6603.}
       Discovered by Hurt et al. (2000) in Sagittarius, the radial velocity of this metal poor, disk globular cluster (Ivanov et 
       al. 2005) is very large, -374 km/s (Borissova et al. 2009). Although the number of young clusters close to this object is 
       not particularly high (2), they are among the closest in the entire pair sample. These clusters are vdBergh 113 (163 pc, 
       32 Myr) and NGC 6561 (277 pc, 8 Myr). They both have very similar radial velocities (-16 km/s vs. -17.93 km/s) but their 
       proper motions are rather different. On the other hand, NGC 6603 (M 24) is also rather close (174 pc, 200 Myr) but its 
       radial velocity is different (21.34 km/s). 2MASS-GC01 also has a small tidal radius, 6.3 pc.        
%
%---------------------------------------------------------------------------------------------------------------------------------
%
%
%---------------------------------------------------------------------------------------------------------------------------------
%

%
%---------------------------------------------------------------------------------------------------------------------------------
%
       \begin{table}[!htb]
         \fontsize{8}{10pt}\selectfont
         \tabcolsep 0.09truecm
         \begin{center}
           \caption{Statistics of pairs within 1 kpc for the NCOVOCC+CPMWGC dataset and objects located within 10 kpc from the 
                    Sun. The separation from the Galactic plane ($Z$-coordinate) in kpc is included in parentheses. 
                   }
           \resizebox{\linewidth}{0.4\linewidth}{ 
           \begin{tabular}{|c|ccccc|}
             \hline  
              Globular    &  Heliocentric  &   Tidal     &  \multicolumn{3}{c|}{Number of pairs} \\ 
              cluster     & distance (kpc) & radius (pc) &  Total  & Young &  Old  \\
             \hline 
              NGC 6397    &    2.3(-0.5)   &   10.6      & 33(6.7$\sigma$) & 8(5.4$\sigma$) & 4(7.5$\sigma$)  \\
              NGC 6121    &    2.2(+0.6)   &   33.2      & 26(5.2$\sigma$) & 9(6.0$\sigma$) & 0               \\
              NGC 6544    &    3.0(-0.1)   &    1.9      & 16(3.2$\sigma$) & 5(3.3$\sigma$) & 2(3.7$\sigma$)  \\
              NGC 6656    &    3.2(-0.4)   &   29.7      &  8(1.6$\sigma$) & 4(2.7$\sigma$) & 1(1.9$\sigma$)  \\
              2MASS-GC01  &    3.6(+0.0)   &    6.3      &  5(1.0$\sigma$) & 2(1.3$\sigma$) & 1(1.9$\sigma$)  \\
              NGC 6838    &    4.0(-0.3)   &   10.4      &  4(0.8$\sigma$) & 1(0.7$\sigma$) & 1(1.9$\sigma$)  \\
              NGC 3201    &    4.9(+0.7)   &   36.1      &  2(0.4$\sigma$) & 0              & 0               \\
              NGC 6366    &    3.5(+1.0)   &   12.1      &  1(0.2$\sigma$) & 1(0.7$\sigma$) & 0               \\
              IC 1276     &    5.4(+0.5)   &   33.9      &  1(0.2$\sigma$) & 1(0.7$\sigma$) & 0               \\
              NGC 6352    &    5.6(-0.7)   &   17.0      &  1(0.2$\sigma$) & 1(0.7$\sigma$) & 0               \\
              NGC 6304    &    5.9(+0.6)   &   22.7      &  1(0.2$\sigma$) & 0              & 0               \\
              Palomar 10  &    5.9(+0.3)   &    5.3      &  1(0.2$\sigma$) & 0              & 1(1.9$\sigma$)  \\
              NGC 4833    &    6.6(-0.9)   &   34.1      &  1(0.2$\sigma$) & 0              & 1(1.9$\sigma$)  \\
             \hline
           \end{tabular}
           }
           \label{candidates}
         \end{center}
       \end{table}
%
%---------------------------------------------------------------------------------------------------------------------------------
%
%
%---------------------------------------------------------------------------------------------------------------------------------
%
       \begin{table}[!htb]
         \tabcolsep 0.05truecm
         \begin{center}
           \caption{Closest pairs. The age corresponds to that of the open cluster.}
           \resizebox{\linewidth}{0.3\linewidth}{ 
           \begin{tabular}{|c|c|cc|}
             \hline  
              Globular    &   Open              & Separation  &  Age   \\ 
              cluster     & cluster             &    (pc)     & (Myr)  \\
             \hline 
              FSR 1767    &  Ruprecht 127       &  36         &   22   \\
              FSR 1767    &  Ruprecht 125       & 112         &  589   \\
              2MASS-GC01  &  van den Bergh 113  & 163         &   32   \\
              2MASS-GC01  &  NGC 6603           & 174         &  200   \\
              FSR 1767    &  Trumpler 28        & 184         &   20   \\
              FSR 1767    &  Collinder 347      & 194         &   12   \\
              NGC 6544    &  ASCC 93            & 236         &   16   \\
              FSR 1767    &  BH 217             & 252         &   45   \\
              FSR 1767    &  BH 202             & 270         &  112   \\
              FSR 1767    &  Teutsch 85         & 275         &  603   \\
              FSR 1767    &  Collinder 351      & 275         &  159   \\
              2MASS-GC01  &  NGC 2561           & 277         &    8   \\
              FSR 1767    &  Hogg 19            & 295         & 1259   \\
              FSR 1767    &  Trumpler 28        & 297         &   12   \\
             \hline
           \end{tabular}
           }
           \label{10pairs}
         \end{center}
       \end{table}
%
%---------------------------------------------------------------------------------------------------------------------------------
%

 \section{Statistical significance}
    A number of facts and trends have been identified during the cluster separation analysis but, how relevant, statistically 
    speaking, are these putative trends? In the following, we estimate the statistical significance of our findings by 
    normalizing the cluster pair counts to the standard deviation ($\sigma$) of the given volume-limited sample. For a given 
    sample, the standard deviation has been computed in the usual way, $\sigma = \sqrt{\sum (n_i - \overline{n}) / n_g}$, where 
    $n_i$ is the number of pairs separated by $<$ 1 kpc for the $i$-th globular cluster, $\overline{n}$ is the average number of 
    pairs, and $n_g$ is the total number of globular clusters in the sample considered. As we can see in Table \ref{candidates}, 
    only 3 globular clusters are statistically significant (following our definition) for the 10 kpc sample and none for the 5 
    kpc one (full sample including young and old clusters). For NCOVOCC+CPMWGC we find: NGC 6397 at 6.7$\sigma$ (33 pairs), NGC 
    6121 (M 4) at 5.2$\sigma$ (26 pairs), and NGC 6544 at 3.2$\sigma$ (16 pairs). NGC 6656 (M 22) is marginally significant at 
    1.6$\sigma$ (8 pairs). The other dataset (NCOVOCC + GGCD+) gives: FSR 1767 at 8.8$\sigma$ (180 pairs), NGC 6397 at 
    1.6$\sigma$ (33 pairs), NGC 6121 (M 4) at 1.3$\sigma$ (26 pairs), NGC 6544 at 1.2$\sigma$ (25 pairs) and NGC 6656 (M 22) at 
    0.4$\sigma$ (8 pairs). For this dataset, FSR 1767 is still marginally significant for the 5 kpc volume-limited sample with 
    180 pairs at 2.9$\sigma$. 

    If we focus on pairs including a young open cluster companion ($<$ 50 Myr), our results remain largely unchanged. For the 
    NCOVOCC+CPMWGC dataset we find: NGC 6397 at 5.4$\sigma$ (8 pairs), NGC 6121 (M 4) at 6.0$\sigma$ (9 pairs) and NGC 6544 at 
    3.3$\sigma$ (5 pairs). NGC 6656 (M 22) is marginally significant at 2.7$\sigma$ (4 pairs). FSR 1767 is still significant for 
    the other dataset at 8.8$\sigma$ with 52 pairs and even for the 5 kpc sample (marginally, 2.8$\sigma$, 52 pairs).

    As for pairs including an old ($>$ 1 Gyr) open cluster, the NCOVOCC+CPMWGC dataset gives NGC 6397 at 7.5$\sigma$ (4 pairs) 
    and NGC 6544 at 3.7$\sigma$ (2 pairs). For the other dataset, FSR 1767 is still significant for old open cluster companions 
    at 8.9$\sigma$ with 24 pairs (for the 5 kpc subsample we find 2.7$\sigma$ with 24 pairs). 

    Table \ref{candidates} includes all the globular clusters with relatively close open cluster companions and summarizes our 
    results for the NCOVOCC+CPMWGC dataset. In total, only 5 globular clusters are statistically significant with FSR 1767 (not 
    included in NCOVOCC+CPMWGC) being a very significant outlier. The closest pairs appear in Table \ref{10pairs}. Among the 
    closest pairs, the majority include a relatively young open cluster. This could be the result of an observational bias 
    because younger open clusters are easier to identify. In the following analysis, we will restrict ourselves to the 
    NCOVOCC+CPMWGC dataset as the presence of the outlier FSR 1767 in the other dataset severely disrupts a clear interpretation 
    as it is the single statistically significant object. On the other hand, the very nature of FSR 1767 is still under debate 
    (see below).   

    If we study the normalized number of open cluster companions ($N_{p}/\sigma$) as a function of the heliocentric distance of 
    the host globular cluster then we realize that the only statistically significant globular clusters are the closest ones. 
    This is just a confirmation of the conclusions obtained from our raw analysis in the previous section. This analysis strongly 
    suggests that, given the significant incompleteness of distant open cluster samples, globular clusters belonging to the disk 
    and bulge populations as well as those from the halo but crossing the disk are routinely surrounded by dozens of open 
    clusters in relative proximity. This interpretation requires a strong but logical assumption: if the surface density of open 
    clusters is uniform and the Solar Neighbourhood is just an average region of the disk (at the Solar Circle), any sphere of 
    radius 1 kpc (or, more properly, any cylinder centred at the disk of radius 1 kpc and height the vertical scale of the disk) 
    centred at the midplane of the Galactic disk, should contain about 400 open clusters of all ages. Therefore, any globular 
    cluster close to the Galactic disk must be surrounded by a relatively large number of open clusters. The alternative to this 
    interpretation is also clear (but wrong): if the statistically significant globular clusters are {\it special}, then the 
    normal scenario is to have very few or no open clusters close to any given globular. This is highly unlikely; the obvious 
    interpretation is that tens of thousands of open clusters remain to be discovered and every globular cluster close to the disk 
    is relatively close to dozens of open clusters that may eventually interact dynamically with it.  

    But how significant is our significance analysis? Are our samples statistically relevant? Most globular clusters are located 
    well above (or below) the Galactic disk. So far, our samples did not account for this effect. There are 48 globular clusters 
    in CPMWGC with heliocentric distance $<$ 10 kpc and $|Z| <$ 1 kpc. The average tidal radius for this subsample is 24.1 pc 
    with a standard deviation of 15.6 pc. If we focus on this best sample we still get statistically significant results: NGC 
    6397 at 6.7$\sigma$ (33 pairs), NGC 6121 (M 4) at 5.2$\sigma$ (26 pairs), NGC 6544 at 3.2$\sigma$ (16 pairs) and NGC 6656 
    (M 22) at 1.6$\sigma$ (8 pairs). 
     
    Van den Bergh (2006) pointed out that open clusters with ages $>$ 1 Gyr appear to form a singular structure that he termed a 
    ``cluster thick disk". Van den Bergh considers that part of the open cluster thick disk consists of objects that were probably 
    captured gravitationally by the main body of the Galaxy. Similar views appear in Gozha et al. (2012). This does 
    not necessarily imply that a fully formed star cluster was captured. NCOVOCC includes 2174 open clusters; out of them, 1629 
    have both distance and age. Within this smaller sample, we find 293 (18~\%) clusters of all ages with separation from 
    the disk $>$ 200 pc. Among younger clusters (age $<$ 100 Myr) the fraction of high altitude objects is nearly 5~\% (26 
    out of 533 clusters). This fraction increases to nearly 50~\% (158 out of 312 clusters) for old clusters (age $>$ 1 
    Gyr). As expected, moving in an inclined orbit increases the survival opportunities of an open cluster. The mechanism 
    presented in Sect. 2 may play a key role on the long-term evolution of open clusters. 
 
    A relatively minor issue can be raised at this stage: the distances to many clusters (both globular and open), in particular 
    the more distant ones, are affected by errors. Hence, these uncertainties in the distance are of the order of the nearest 
    neighbour distance applied in our analysis and they may put our conclusions in jeopardy. As a matter of fact, this concern is 
    unfounded as most of our samples are volume limited. A Monte Carlo simulation of 10$^5$ artificial data sets using original 
    cluster samples but altered by random errors up to 30~\% in distance (see above) gives similar numbers of statistically 
    significant globular clusters with nearly the same number of pairs per cluster but, as expected, the actual identity of the 
    statistically significant globular clusters and their closest open cluster companions changes.

 \section{Monte Carlo approach}
\begin{figure*}[!htb]
      \centering
        \includegraphics[width=\textwidth]{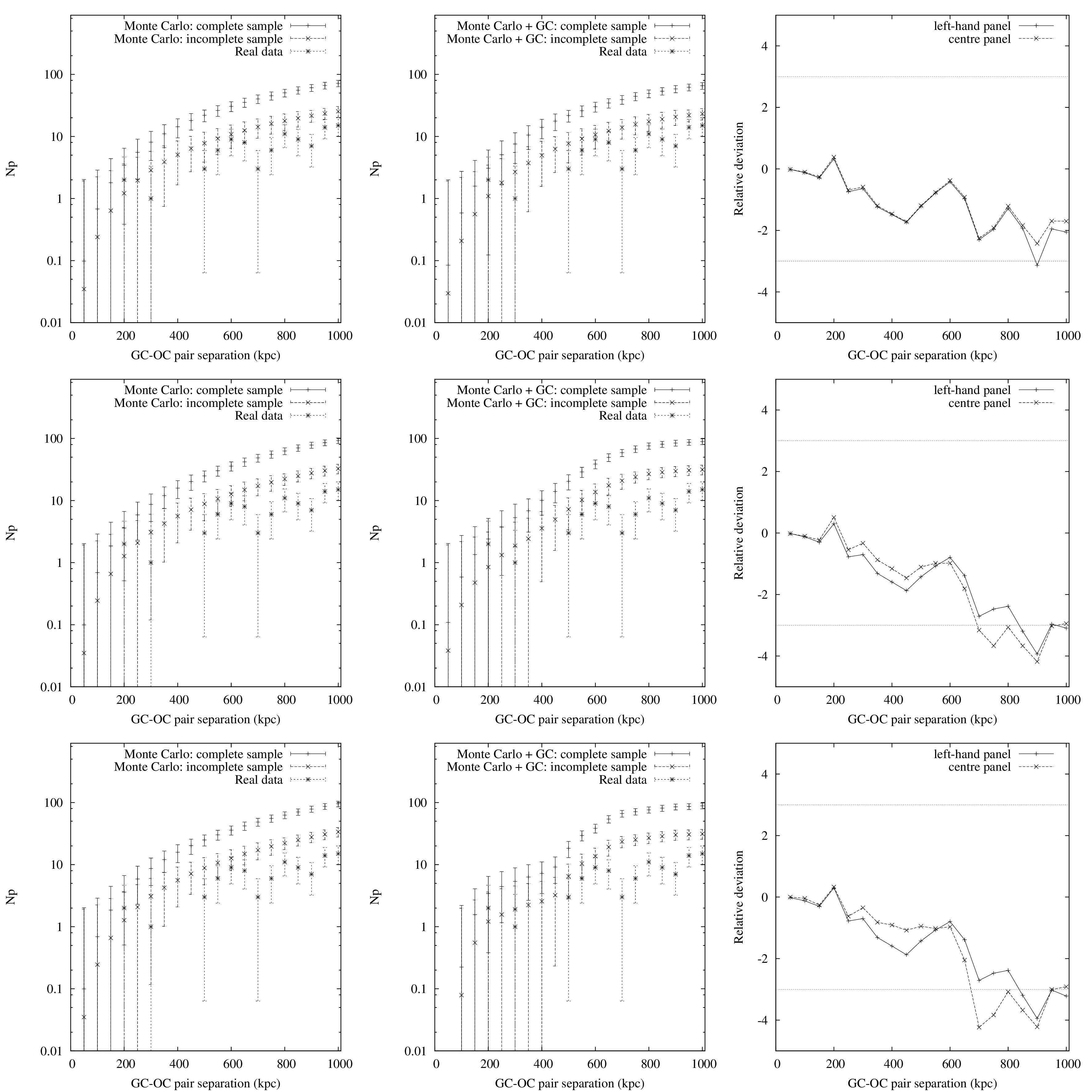}
        \caption{Number of GC-OC pairs as a function of their separation. This figure is formally equivalent to Fig. 
                 \ref{evidence} (bottom panels) but it focuses on globular and open clusters located within 3 kpc from the Sun 
                 and also with intercluster distances $<$ 1 kpc. The synthetic data used here have been obtained using the Monte 
                 Carlo approach described in the text. The left-hand panel shows randomly generated positions for both globular 
                 and open clusters. The central panel includes the actual coordinates of the globular clusters NGC 6121, NGC 6397 
                 and NGC 6544 and randomly generated positions for the open clusters. The right-hand panel shows the relative 
                 differences in units of the standard deviation between the number of pairs observed and the one obtained using 
                 the Monte Carlo-type experiment assuming incompleteness. If instead of assuming a thickness of 2 kpc for the 
                 open cluster population (top three panels) we consider 325 pc, we obtain the results in the middle three panels; 
                 if the thickness is just 100 pc we obtain the three panels at the bottom.}
    \label{monte}
  \end{figure*}

    Our previous analyses indicate that globular clusters traveling across the disk are surrounded by an ensemble of open 
    clusters but these open clusters are not found at uniformly distributed separations or following a linear or quadratic spread 
    but preferentially at relatively large separations ($s \gg r_t$, where $r_t$ is the tidal radius of the globular cluster). In 
    the following, we will use Monte Carlo techniques to try to confirm that the gap found is real, likely the result of cluster 
    harassment, and not an artefact induced by the lack of completeness of the observational samples. Let us consider an 
    environment similar to the Solar Neighbourhood within 3 kpc from the Sun: we neglect the special case of FSR 1767 and we also 
    consider the completeness correction for open clusters as outlined in the previous sections (cluster number proportional to 
    the square of the distance and a total population of 400 open clusters within 1 kpc from the Sun). Therefore and in theory, 
    we have three globular clusters (NGC 6121, NGC 6397 and NGC 6544) and 3600 open clusters. Let us remind the reader that, from 
    NCOVOCC, we have 389 open clusters within 1 kpc from the Sun and we estimate that this sample is nearly 100~\% complete. 
    In our first Monte Carlo-type calculation we consider this scenario: 3 globular clusters and 3600 open clusters with $X$ and 
    $Y$ coordinates randomly generated in the range [-3, 3] kpc and $Z$ in the range [-1, 1] kpc, all within a cylinder of radius 
    3 kpc and thickness 2 kpc to study the intercluster separation as we did in Fig. \ref{evidence}, bottom panels. This 
    corresponds to results labelled ``Monte Carlo: complete sample" in Fig. \ref{monte} (top left panel) and includes 3 globular 
    clusters and 3600 open clusters, a complete sample of both globular and open clusters. On the other hand, there are 1274 open 
    clusters within 3 kpc from the Sun in NCOVOCC (this sample is complete at the 35~\% level, as pointed out above) and we 
    account for this fact in our second Monte Carlo-type calculation: now 3 globular clusters and 1274 open clusters with 
    randomly generated coordinates as described above. This corresponds to results labelled ``Monte Carlo: incomplete sample" in 
    Fig. \ref{monte} (top left panel). The central panel in Fig. \ref{monte} uses randomly generated positions for the open 
    clusters but includes the actual coordinates of NGC 6121, NGC 6397 and NGC 6544. Data plotted in both panels are compatible; 
    therefore, the actual values of the coordinates of the globular clusters are unimportant here. Error bars display the 
    1$\sigma$ Poisson errors as described in Sect. 5.2. The right-hand panel shows the relative differences in units of the 
    standard deviation between the number of pairs observed and the one obtained using the Monte Carlo-type experiment assuming 
    incompleteness. We have performed 2$\times10^6$ experiments for each simulated case in order to produce statistically 
    meaningful results. The randomly generated distributions implicitly assume that both globular and open clusters are 
    non-interacting populations, i.e. their positions are uncorrelated. Results obtained for the more realistic vertical scale 
    heights of 325 pc or 100 pc are displayed on the middle and bottom panels, respectively. For the most realistic models, 
    statistically significant deviations from the non-collisional scenario are found. These deviations are consistent with those
    observed in Fig. \ref{evidence} even if the methodology used is different.
          
%
%---------------------------------------------------------------------------------------------------------------------------------
%
%
%---------------------------------------------------------------------------------------------------------------------------------
%
    Our simulated incomplete samples cannot reproduce (within the error limits) the observational data across the entire 
    separation range. They clearly fail to reproduce the observed scarcity of closer pairs, particularly for the separation range
    650-1000 pc. No pairs closer than 150 pc are expected, statistically; therefore, the lack of very close pairs observed in Fig. 
    \ref{pairs} and initially attributed to tidal effects can also be explained on purely statistical grounds. Results are 
    similar when using real globular cluster coordinates (central panels in Fig. \ref{monte}). There is a statistically 
    significant deficit of close pairs with respect to the non-collisional scenario envisioned in the Monte Carlo calculations. 
    In summary, the assumed non-collisional scenario is not well supported by our results. If interactions between globular and 
    open clusters were negligible, a larger number of close pairs should be observed. Claiming incompleteness of the samples 
    cannot explain the observed deficit; the alternative explanation appears obvious. Globular and open clusters do interact but 
    destructively, most of the time. On the other hand and if FSR 1767 is a real globular cluster, it clearly matches what is 
    expected of a globular cluster starting to cross the disk. It is surrounded by a large number of open clusters, consistent 
    (in principle) with the one predicted by our Monte Carlo calculations. FSR 1767 has 180 open clusters within 1 kpc from its 
    centre; the Monte Carlo calculations assuming complete samples predict 232. But FSR 1767 is located 1.5 kpc from the Sun; 
    therefore, the set of open clusters surrounding the object is not expected to be complete and additional clusters likely 
    remain to be discovered in that area. Assuming that the open cluster sample located within 2 kpc of the Sun is about 70~\% 
    complete then the expected number could be as high as 257; on the other hand, if the sample is 50~\% complete then 
    the expected number is 360, that is 8.4$\sigma$ the Monte Carlo-predicted value. Therefore, FSR 1767 appears to be inducing 
    formation of open clusters if completeness is considered. In conclusion, FSR 1767 may be surrounded by a number of open 
    clusters significantly larger than the one predicted by a simple non-collisional scenario. Within this analysis, FSR 1767 
    emerges as an archetypal globular cluster crossing the disk and surrounded by its ephemeral cloud of open clusters and likely 
    inducing star formation in its neighbourhood. 
%
%---------------------------------------------------------------------------------------------------------------------------------
%
    \begin{figure*}
      \centering
        \includegraphics[width=\textwidth]{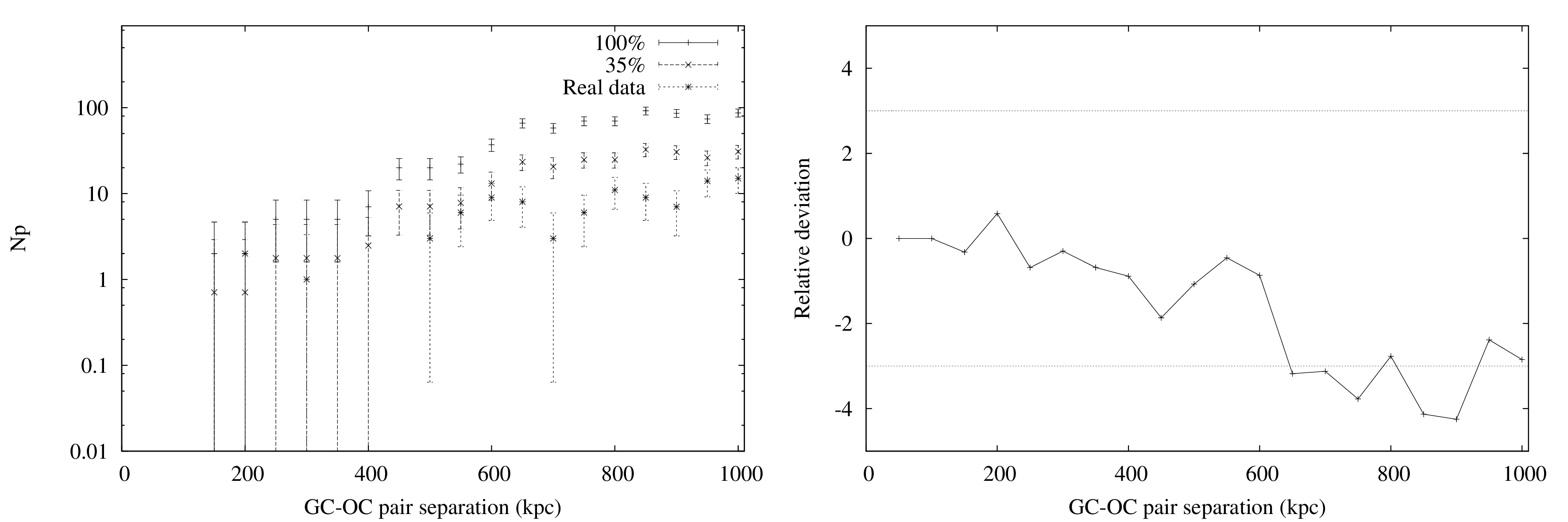}
        \caption{Number of GC-OC pairs as a function of their separation for the replicated fractal distribution. The artificial 
                 cluster sample has been obtained as described in the text and it has fractal structure. The number of pairs 
                 displayed are associated to samples with different degree of completeness, 100 and 35~\%. Real data as in 
                 Fig. \ref{monte} are also plotted. The right-hand panel shows the relative differences in units of the standard 
                 deviation between the number of pairs observed and the one obtained using the model with a degree of 
                 completeness of 35~\%.} 
      \label{clone}
    \end{figure*}
%
%---------------------------------------------------------------------------------------------------------------------------------
%

    The Monte Carlo calculation presented above makes a major implicit assumption: clusters are distributed uniformly in the disk. 
    Unfortunately, the actual spatial distribution of open clusters in the Milky Way disk is not that simple. The spatial 
    distribution of young open clusters in the Solar Neighbourhood exhibits multifractal structure (de la Fuente Marcos and de la 
    Fuente Marcos 2006). For obvious reasons, the majority of open clusters in our samples are young. Besides, the fractal 
    dimension is time-dependent, increasing over time. Even if the positions of the clusters are organized following a fractal 
    pattern, it is not clear which fractal model reproduces the overall spatial distribution better. There is however an 
    alternative to using a particular fractal model, we can use the actual observed distribution (which is fractal by itself). 
    The sample of open clusters located within 1 kpc from the Sun is almost 100~\% complete and it has not been perturbed 
    by globular clusters during the last 20 Myr at least. In the following, we replicate the spatial distribution of open 
    clusters found in a square of side 2 kpc centred on the Sun by shifting their coordinates to generate a grid of 9 equal 
    squares and then we restrict the artificial sample created to those objects with $X$ and $Y$ coordinates in the range [-3, 3] 
    kpc and $Z$ in the range [-1, 1] kpc all within a cylinder of radius 3 kpc and thickness 2 kpc. These are ``real" coordinates 
    so they take into account that most open clusters are found within 200 pc from the Galactic plane. Then we use the 
    artificially generated (complete) fractal sample to study the intercluster distance separation. This approach has multiple 
    advantages over the previous one: we use a realistic fractal distribution of positions, the template sample (used in the 
    replication process) was not under the direct effects of neighbouring globular clusters, and the sample is complete. Here, we 
    are implicitly assuming that every location within the Solar Circle is equivalent which is a very reasonable assumption. Our 
    analysis includes the actual coordinates of NGC 6121, NGC 6397 and NGC 6544. The calculation has been repeated by randomly 
    removing a fraction of the clusters to simulate different levels of completeness so the number of pairs displayed are 
    associated to samples with different degree of completeness although the results focus on 100 and 35~\%. As in the 
    previous simulations, 2$\times10^6$ trials for each sample have been computed in order to provide statistical significance to 
    our results. Figure \ref{clone} confirms again the scarcity of close pairs with respect to a model in which globular clusters 
    have no effects on the spatial distribution of open clusters. The results are consistent with those in Figs. \ref{evidence} 
    and \ref{monte}. Therefore, we can conclude that the average number of open clusters close to a globular cluster in the 
    Solar Neighbourhood is significantly below the expected value in a purely non-collisional scenario: globular and open 
    clusters are indeed interacting populations and globular clusters do affect the environment in which they move. 

    It may be argued that the statistically significant deviations appear only for separations $>$ 500 pc when the tidal effects
    are expected to be more important at smaller separations. As pointed out before, statistical fluctuations are responsible for
    that; the ratio between the number of pairs and the standard deviation is only $>$ 1 for $N_p > 3$ and $> 3$ for $N_p > 15$.
    Therefore, if the actual number of pairs is 0 and the one predicted by the non-collisional model is 10, the relative 
    deviation in terms of $\sigma$ is $<$ 3 although the actual difference is far from negligible. These fluctuations actually
    mask the mismatch between the real data and those from the non-collisional model at smaller separations making real data 
    apparently compatible with the non-collisional scenario. For a much larger sample, instead of a few thousand clusters, 
    statistically significant differences should be observed across the entire range of separations. Our results for simulated 
    samples are fully consistent with our previous theoretical analysis in Sect. 2: in the Milky Way disk, interactions between 
    globular and open clusters are far more frequent than commonly thought and the outcome is usually the destruction of the open 
    clusters involved. 

 \section{FSR 1767: a statistically significant outlier} 
    In the statistical analysis of the dynamical connection between globular and open clusters, FSR 1767 clearly stands out in 
    the sense that it has the most open clusters in its vicinity. Its long list of neighbouring young open clusters includes 
    Ruprecht 127 (36 pc, 22 Myr), Trumpler 28 (184 pc, 19 Myr), Collinder 347 (194 pc, 12 Myr), BH 217 (252 pc, 45 Myr), Trumpler 
    27 (297 pc, 12 Myr), BH 200 (304 pc, 22 Myr), NGC 6396 (310 pc, 32 Myr), NGC 6231 (346 pc, 7 Myr) and NGC 6530 (371 pc, 7 
    Myr). These objects are located less than 400 pc from the cluster, separations and ages are provided in parentheses. Looking 
    a little farther away we find Hogg 22, Bochum 13, Collinder 367, Trumpler 24, NGC 6531, ASC 88, ASC 85, NGC 6268, NGC 6242, 
    NGC 6193 and Pismis 24. NGC 6193 is the core of the Ara OB 1a stellar association and is related to the RCW 108 complex. Ara 
    OB 1a is an intriguing star-forming region that is considered as one of the best examples of triggered star formation even if 
    the identification of the actual triggering mechanism remains elusive (Wolk et al. 2008). For an assumed 
    cluster velocity of 100-200 km/s, the ballistic time-of-flight from Ara OB 1a to the present location of FSR 1767 is 5-2 Myr 
    which is formally consistent with the age of the star-forming complex if the trigger was FSR 1767. Baume et al. (2011) have 
    concluded that NGC 6193 is a very young open cluster (1-5 Myr) located in the Sagittarius-Carina Galactic arm and within the 
    Ara OB1 complex. In general, the previous list of objects exhibit similar proper motions which may suggest that all of them 
    were formed nearly at the same time perhaps by the passing FSR 1767. Then, we find Trumpler 24 (413 pc, 8 Myr, (-0.45, -0.82) 
    mas/yr), NGC 6242 (417 pc, 41 Myr, (0.33, 0.10) mas/yr), ASCC 85 (424 pc, 26 Myr, (-0.43, -4.01) mas/yr), Bochum 13 (431 pc, 
    7 Myr, (-2.17, -2.84) mas/yr), Hogg 22 (436 pc, 6 Myr, (-1.46, -4.55) mas/yr), ASCC 88 (436 pc, 15 Myr, (2.89, -2) mas/yr), 
    NGC 6268 (451 pc, 40 Myr, (0.13, -0.85) mas/yr), Collinder 367 (453 pc, 7 Myr, (1.16, -2.03) mas/yr), NGC 6531 (462 pc, 12 
    Myr, (2.28, -3.05) mas/yr) and NGC 6193 (502 pc, 6 Myr, (0.23, -4.48) mas/yr). Proper motions are affected by very large 
    errors (see NCOVOCC or WEBDA for details) but the kinematic evidence is rather encouraging. For this unusual object we find 
    both high concentration of young open clusters in its immediate neighbourhood and signs of possible tidal perturbation. In 
    principle, a distorted density profile supports the case of it having triggered formation of open clusters but as a globular 
    cluster passes through the Galactic disk it is also expected to undergo tidal distortion from the gravitational potential of 
    the entire disk. In order to arrive to solid conclusions not conjectures, the hypothetical tidal tails of this object should 
    be studied but that requires extensive imaging that it is not yet available.

    On the other hand, the list of close older neighbours includes NGC 6253 (460 pc, 5 Gyr), ESO 282-26 (510 pc, 1.3 Gyr), NGC 
    6698 (551 pc, 1.9 Gyr), ESO 397-01 (579 pc, 1.3 Gyr), Lynga 8 (618 pc, 2 Gyr), Ruprecht 138 (628 pc, 2.0 Gyr), NGC 5998 (635 
    pc, 1.6 Gyr), Ruprecht 171 (650 pc, 3.2 Gyr), ESO 139-54 (662 pc, 1.3 Gyr), IC 4651 (671 pc, 1.1 Gyr) and NGC 6208 (686 pc, 
    1.2 Gyr). But and following Carraro and Bensby (2009), do we have any common kinematic and/or chemical signature? Its radial 
    velocity has not yet been determined but proper motions, (2.81$\pm$2.85, -8.78$\pm$2.82) mas/yr, were computed by Bonatto et 
    al. (2007). There, an age of 12 Gyr and a metallicity of [Fe/H] = -1.2 are suggested for FSR 1767. Most of the old open 
    clusters cited above do not have kinematic or chemical information available from WEBDA or NCOVOCC and none of the few 
    available are consistent with the properties of FSR 1767. Therefore and even if the fraction of neighbouring old open clusters 
    is statistically significant, no other available data support a possible dynamical connection between this globular cluster 
    and any of the objects in its long list of neighbouring old open clusters.

    But, is FSR 1767 a real globular cluster? Unfortunately, the very nature of FSR 1767 as a globular cluster is still under 
    debate. Located in Scorpius, this Palomar-like object was discovered by Froebrich et al. (2007) and further 
    studied by Bonatto et al. (2007). It is one of the closest globular clusters (1.5 kpc from the Sun), 50 pc below the Galactic 
    plane. It has a very small core of 0.24 pc and a tidal radius of 3.1 pc (Bonatto et al. 2007). If this cluster is passing 
    through the Galactic disk, it may have been stripped of a large fraction of its outermost members leaving just the stronger 
    gravitationally bound core. The cluster appears very distorted although no formal calculation of its ellipticity has been 
    attempted. The object has been re-analyzed using new NTT observations by Froebrich et al. (2009) to conclude that it is not a 
    star cluster but an observational artefact caused by differential extinction. However, this negative conclusion has been 
    disputed: Bonatto et al. (2009) combined old and new data to show that the properties of the object are consistent with those 
    of a globular cluster. Their main argument against Froebrich et al. (2009) is that their calibration of the FSR 1767 data is 
    erroneous. However, for the other clusters in the same dataset results from both groups are fully consistent and the original 
    SOFI/NTT data were not re-analyzed in Bonatto et al. (2009). For them, FSR 1767 is a nearby low-luminosity, relatively 
    metal-poor globular cluster projected against the bulge. They also found a peculiar compact stellar group in the central 
    parts of the cluster that they interpret as a possible detached post-collapse core. However, {\it Chandra} observations 
    looking for a putative X-ray source population in the field of FSR 1767 using the new coordinates provided by Bonatto et al. 
    (2009) have proven negative. No X-ray sources within 4 arcminutes of the new coordinates have been detected. This finding 
    rules out the presence of any quiescent X-ray binaries. Fainter objects such as millisecond radio pulsars or cataclysmic 
    variables were below the detection limit of the {\it Chandra} observations, therefore their presence cannot be ruled 
    out\footnote{Information provided by E. M. Cackett (November 2009)}. Detection of clustered X-ray sources is customarily 
    regarded as solid confirmation of the globular cluster nature of an old stellar ensemble. 

 \section{Our results in context}
    The topic of how globular clusters are affected by the Galactic disk has been investigated for decades but the study of their 
    impact on the disk structures is relatively new. To put our current results into perspective, cluster shocking during passage 
    through the Galactic disk has long been considered a main threat to the long-term survivability of globular clusters 
    (Ostriker et al. 1972) but the reaction of the disk to that passage was not studied or even mentioned until much later. In a 
    little-known paper, Brosche et al. (1991) stated that ``Every transit of a globular cluster through the interstellar matter in 
    the galactic plane will, in principle, affect this material. There is slight hope that such perturbing effects may be 
    noticeable even today." For these authors, the expected signature would take the form of a remnant ``shot hole" imprinted in 
    the interstellar matter. High-velocity clouds can also generate a cavity when they collide with the Galactic disk 
    (Tenorio-Tagle 1981; Tenorio-Tagle et al. 1986). The capture of field stars by passing globular clusters has been studied 
    mainly by Bica et al. (1997) and Mieske and Baumgardt (2007), the latter found it not very efficient. On both theoretical and 
    observational grounds, present-day, tidally triggered star formation in the Milky Way disk has been suggested as a plausible 
    secondary mechanism capable of forming open clusters (Wallin et al. 1996; Levy 2000; de la Fuente Marcos and de la Fuente 
    Marcos 2008b; Vande Putte and Cropper 2009). The implied tidal encounter paradigm involves interactions between passing 
    globular clusters (and perhaps dwarf galaxies too) and giant molecular clouds. A version of this mechanism based on Toomre's 
    parameter was first applied to globular clusters crossing along paths perpendicular to the Galactic plane by Wallin et al. 
    (1996) although their theoretical arguments may not be valid for clusters moving in paths not strictly normal to the disk (de 
    la Fuente Marcos and de la Fuente Marcos 2008b), i.e. globular clusters in non-polar orbits. Nevertheless, hydrodynamic 
    simulations carried out by Levy (2000) and Vande Putte and Cropper (2009) lent further support to the early results obtained 
    by Wallin et al. (1996). De la Fuente Marcos and de la Fuente Marcos (2008b) used a different version of the tidal encounter 
    paradigm in an attempt to explain present-day star formation at high Galactic altitude. In this case, the analysis was based 
    on the Jeans' criterion under the distant-tide and impulse approximations (Binney and Tremaine 2008). This new version of the 
    paradigm does not assume any particular inclination for the orbit of the perturbing globular cluster. The topic has been 
    brought again to the attention of the astronomical community by Salerno et al. (2009), suggesting that $\omega$ Centauri was 
    involved in the formation of the young massive open clusters Stephenson 2 and BDSB 122. In their paper, it is claimed that 
    current evidence is consistent with globular clusters being additional progenitors of open clusters; in particular, the most 
    massive ones.  

    Vande Putte and Cropper (2009) have identified three globular clusters as strong candidates to being responsible for tidally 
    induced star formation. Two of them, NGC 6397 and NGC 3201, have young neighbour clusters following the criteria used in the 
    present work although only the case of NGC 6397 appears to be statistically significant. Vande Putte and Cropper (2009) 
    consider that some of the stellar associations observed close to NGC 6397 are remnants of a recent interaction. In our 
    analysis, NGC 6397 appears as statistically significant and therefore consistent with their interpretation. NGC 3201 is a 
    globular cluster in Vela with very low central stellar concentration and ellipticity 0.12. It has only two open clusters 
    within 1 kpc. The third candidate identified in Vande Putte and Cropper (2009), NGC 6838 (M 71), has one young neighbour but 
    three other clusters are close to it. Located in Sagitta, it is one of the globular clusters exhibiting abundance variations 
    (Alves-Brito et al. 2008). The young open cluster NGC 6231 is considered by Vande Putte and Cropper (2009) as not related to 
    any globular cluster-induced star formation episode but our results do not support this conclusion as it is one of the close 
    neighbours of FSR 1767. In general, our results appear to be consistent with those from other authors. Vande Putte et al. 
    (2009) have also suggested that passing globular clusters may contribute to the Galactic disk heating. Our analysis of the 
    results obtained from Eq. (\ref{timescale}) also favours an scenario in which globular clusters contribute towards the 
    vertical heating of disk objects. Vande Putte et al. (2010) have analyzed the orbits of 481 open clusters to conclude that 
    three of them (NGC 1817, NGC 6791 and NGC 7044) may have formed as a result of the impact of a globular cluster on the disk. 
    Our present results do not support their conclusions. 

 \section{Conclusions}
    As a manner of summary and in the Milky Way, theory and observations consistently indicate that globular and open clusters
    certainly are interacting populations. Globular and open clusters have traditionally been considered as completely unrelated
    only because no globular clusters are present within 1 kpc from the Sun and our data on open clusters are very incomplete; in
    contrast, the emerging picture from our analysis is quite the opposite. At the Solar Circle, there is an obvious scarcity of 
    close globular-open cluster pairs and the likely explanation is linked to passing globular clusters that can tidally disrupt 
    small open clusters in a short time-scale. In Sect. 1, we pointed out that it is conventionally assumed that globular and open 
    clusters never interact, they have virtually no connections. This paper has explored the reliability of this assumption and 
    has found reasonable statistical evidence to discard it. Globular and open clusters are indeed connected, this connection is 
    of dynamical nature, and it may include several distinct manifestations: cluster-cluster interactions or cluster harassment, 
    induced open cluster formation by globulars and possible association of old open clusters to accreted globulars. We have 
    investigated the statistical strength of these manifestations further using samples from several databases, recent results 
    from the literature and Monte Carlo-type calculations. FSR 1767, if real, could be the archetypal example of this scenario in 
    which globular and open clusters interact at will. In this respect, another object deserving further study is 2MASS-GC01. Both 
    objects appear to be rather unique and better observations of them are urgently needed. 

    Both theory and observations indicate that repeated weak and distant encounters between open clusters and traveling globular 
    clusters may contribute towards tidal truncation, in the case of globular clusters, and accelerated disruption (in just a few 
    Myr), in the case of open clusters. The strength of this process, tidal destruction of open clusters by passing globular 
    clusters, seems to be quite significant but, so far, customarily ignored. This may well be the evolutionary path for the 
    cluster pair FSR 1767/Ruprecht 127, with a separation of just 36 pc. The actual existence of Ruprecht 127 is however debatable. 
    McSwain and Gies (2005) consider this object an asterism: a chance physical alignment of several bright stars. However, 
    Irrgang et al. (2010) suggested Ruprecht 127 as one of the possible origins of the hyper-runaway candidate HIP 60350. 
    Regarding the issue of present-day open cluster formation induced by the passage of globular clusters, the amount of induced 
    star formation seems to be modest as expected of a secondary mechanism. The fraction of globular clusters currently involved 
    in these events appears to be relatively small although if we restrict our analysis to globular clusters located within 3 kpc 
    from the Sun, the entire sample appears to have experienced relatively close encounters with young open clusters within the 
    last 10 Myr and likely all of them may have been involved in tidally induced star formation events. With the exception of 
    FSR 1767, dynamical association between old open clusters and globular clusters is not supported by currently available data.  

    In this introductory paper, we have presented robust theoretical and statistical evidence on the existence of gravitational 
    interactions between globular and open clusters and their effects on both cluster populations but yet many other interesting
    related topics have been left outside of our study. Areas not covered here include a detailed dynamical analysis of such 
    interactions, their effects on the chemo-dynamical evolution of the Galactic disk, the importance (in absolute terms) of the
    shock-induced star and cluster formation, the possibility of generating a thick disk of stars as a result of these 
    interactions, the potential role of the open cluster thick disk as tracer of the chemical and age evolution of the disk, the
    capture of stars from the tidally disrupted open clusters by the interacting globular clusters, their connection with the
    observed stellar streams, among others. Unfortunately, the serious study of many of these topics, although certainly desirable, 
    requires high quality observational data sets that are not yet available. Future surveys, in our Galaxy and others, may 
    determine the full extent of the impact of the processes discussed here on the evolution of galactic disks.

 \acknowledgments
    The authors would like to thank the anonymous referee for his/her very constructive report and helpful suggestions, and Edward 
    Cackett for providing his unpublished {\it Chandra} results on FSR 1767. RFM and CFM acknowledge partial support by the 
    Spanish `Comunidad de Madrid' under grant CAM S2009/ESP-1496. This research has made use of NASA's Astrophysics Data System 
    Bibliographic Services and the ASTRO-PH e-print server. In preparation of this paper, we made use of the WEBDA database 
    operated at the Institute of Astronomy of the University of Vienna, Austria. This work also has made extensive use of the 
    SIMBAD database and the VizieR catalogue access tool, operated at the CDS, Strasbourg, France.

\end{document}